\documentclass{article}
\usepackage[margin=1.25in]{geometry}
\usepackage{amsmath}
\usepackage{graphicx}
\usepackage{amsthm}
\usepackage{amssymb}
\usepackage{lineno}
\usepackage[caption=false]{subfig}
\usepackage{tikz}
\usepackage{hyperref}
\usepackage{enumerate}
\usepackage{rotating}
\usepackage[authoryear,sort&compress]{natbib}
\DeclareMathOperator{\erf}{erf}
\usetikzlibrary{shapes.geometric, arrows, decorations.pathreplacing, shapes.arrows}
\tikzstyle{stage}=[circle,text centered,draw=black,fill=white]
\tikzstyle{arrow} = [thick,->,>=stealth]
\tikzstyle{stage2}=[circle,text centered,draw=black,fill=white, minimum size=1cm]

\title{Calculating the timing and probability of arrival for sea lice dispersing between salmon farms}

\author{Peter D. Harrington\thanks{Department of Mathematical and Statistical Sciences, University of Alberta, Edmonton, Canada. \newline (\href{mailto:harringt@ualberta.ca}{harringt@ualberta.ca},  \href{mailto:mark.lewis@ualberta.ca}{mark.lewis@ualberta.ca})}\, ,  Danielle L. Cantrell\thanks{California Department of Fish and Wildlife, Marine Region's Fisheries Analytics Project, 20 Lower Ragsdale Drive, Suite 100, CA 93940}, Michael G. G. Foreman\thanks{Institute of Ocean Sciences, Fisheries and Oceans Canada}, \\ Ming Guo\footnotemark[3]  and  Mark A. Lewis\footnotemark[2] \thanks{Department of Biological Sciences, University of Alberta, Edmonton, Canada.}}

\begin{document}

\maketitle

\begin{abstract}

Sea lice are a threat to the health of both wild and farmed salmon and an economic burden for salmon farms. With a free living larval stage, sea lice can disperse tens of kilometers in the ocean between salmon farms, leading to connected sea lice populations that are difficult to control in isolation. In this paper we develop a simple analytical model for the dispersal of sea lice between two salmon farms. From the model we calculate the arrival time distribution of sea lice dispersing between farms, as well as the level of cross-infection of sea lice.  We also use numerical flows from a hydrodynamic model, coupled with a particle tracking model, to directly calculate the arrival time of sea lice dispersing between two farms in the Broughton Archipelago, BC, in order to fit our analytical model and find realistic parameter estimates. Using the parametrized analytical model we show that there is often an intermediate inter-farm spacing that maximizes the level of cross infection between farms, and that increased temperatures will lead to increased levels of cross infection.

% for the arrival time distribution of sea lice dispersing between two different salmon farms, from which we can determine the level of cross-infection between farms.

\end{abstract}

\section{Introduction}

Marine populations are often connected over large distances due to larval dispersal. Once thought to be open populations with continuous exchanges of larvae, it is now understood that many marine populations depend directly on the degree of larval exchange between population patches, and that these connected patches act as metapopulations \citep{Cowen2000, Cowen2006, Cowen2009}. The degree of connectivity between habitat patches in a metapopulation is a function of many variables, including the strength of the ocean currents on which the larvae depend to disperse and the environmental conditions of the ocean which impact biological processes such as maturation and survival. Research into larval dispersal in marine metapopulations has led to a greater understanding of the population dynamics of corals \citep{MayorgaAdame2017}, coral reef fish \citep{Jones2009} and sea turtles \citep{Robson2017}, as well as the efficacy of Marine Protected Areas \citep{Fox2016, Botsford2009}. It has also been used to determine the level of sea lice dispersal between salmon farms and the effect of coordinated treatment plans in salmon farming regions \citep{Samsing2017, Cantrell2018, Adams2015, Kragesteen2018}.  
 
% depends directly on the ocean currents by which the larvae disperse and the strength of these ocean currents and the ability of larvae to disperse are dependent on the level of diffusive and advective movement between patches, as well as the temperature and salinity of the ocean.

\begin{figure}
\centering
\begin{tikzpicture}
\node (freeliving) [draw=none, fill=none] {Free-living};
\node (cop) [draw=black,fill=none, below of=freeliving, yshift=-1cm]  {Copepodite};
\node (nauplius) [draw=black,fill=none, below of=cop, yshift=-0.25cm]  {Nauplius};
\node (attached) [draw=none, fill=none, right of=freeliving, xshift=2cm] {Attached};
\node (chalimus) [draw=black, fill=none, below of=attached, yshift=-0.35cm] {Chalimus};
\node (preadult) [draw=black, fill=none, below of=chalimus, yshift=-0.25cm] {Pre-adult};
\node (adult) [draw=black, fill=none, below of=preadult, yshift=-0.25cm] {Adult};
\node (inv1) [draw=none, fill=none, right of=freeliving, xshift=0.6cm, yshift=-0.4cm] {};
\node (inv2) [draw=none, fill=none, below of=inv1, yshift=-3.4cm] {};
\node (inv3) [draw=none, fill=none, above of=inv1] {};
\node (inv4) [draw=none, fill=none, below of=inv2, yshift=0.1cm] {};
\draw [loosely dotted, thick, line cap=round] (inv3)--(inv4);
\draw [arrow] (cop) to [out=90, in=180] (inv1);
\draw [arrow] (inv1) to [out=0, in=90] (chalimus);
\draw [arrow] (chalimus)--(preadult);
\draw [arrow] (preadult)--(adult);
\draw [arrow] (adult) to [out=270, in=0] (inv2);
\draw [arrow] (inv2) to [out=180, in=270] (nauplius);
\draw [arrow] (nauplius)--(cop);
\end{tikzpicture}
\caption{A simplified schematic of the life cycle of the sea louse, \textit{Lepeophtheirus salmonis}. The attached stages live on wild or farmed salmon and the free-living stages disperse in the water column. Larvae must mature through the nauplius stage into the copepodite stage before they are able to attach to a salmonid host. \label{fig:lifecycle}}
\end{figure}
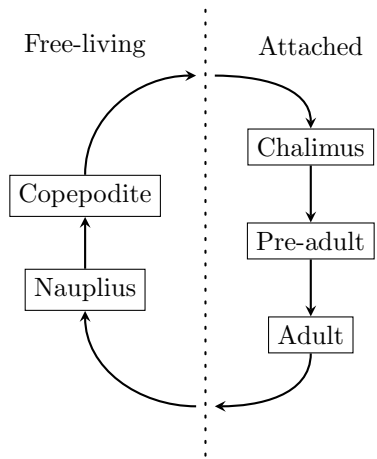

Sea lice (\textit{Lepeophtheirus salmonis}) are parasitic marine copepods that feed on the epidermal tissues, muscle, and blood of salmon \citep{Costello2006}. A free living nauplius stage allows sea lice to disperse tens of kilometers in the ocean while developing into infectious copepodites that can attach to their salmonid hosts, on which sea lice complete the remainder of their life cycle (see Figure \ref{fig:lifecycle}) \citep{Amundrud2009, Stucchi2011}. High infestation levels on adult salmon have been shown to lead to mortality and morbidity \citep{Pike1999}, and lesions and stress from infestations make adult salmon susceptible to secondary infections, which have led to large economic consequences for the salmon farming industry \citep{Costello2009}. On wild juvenile salmon, infestation with sea lice can lead to mortality \citep{Krkosek2007} or other physiological \citep{Brauner2012} and behavioural effects \citep{Krkosek2011a, Godwin2015}. In near coastal areas, elevated levels of sea lice from salmon farms have been detected on juvenile salmon up to tens of kilometers away and these high levels of infection have contributed to population level declines in pink salmon \citep{Krkosek2007, Krkosek2005, Krkosek2006, Peacock2020}.

%Salmon farms in near coastal areas act as a source of sea lice for juvenile wild salmon, which otherwise would not be exposed to sea lice until they mature to adulthood \citep{Krkosek2005, Krkosek2006, Peacock2020}. Elevated levels of sea lice from salmon farms have been detected on juvenile salmon up to tens of kilometers away and these high levels of infection have led to population level declines in pink salmon \citep{Krkosek2007}. To prevent high levels of sea lice on wild salmon, salmon farms in British Columbia are now required to treat once a threshold of three adult motile lice is reached on a farm \citep{FisheriesandOceansCanada2015}, leading to a cessation of the population decline in pink salmon \citep{Peacock2013}. However, even with these new regulations there have been sea lice outbreaks on wild salmon \citep{Bateman2016}, potentially caused by high temperatures and poorly timed treatments on farms, as well as reports of farms underreporting sea louse counts \citep{Godwin2021}.

\begin{figure}
\centering
\subfloat[]{
\includegraphics[width=11cm]{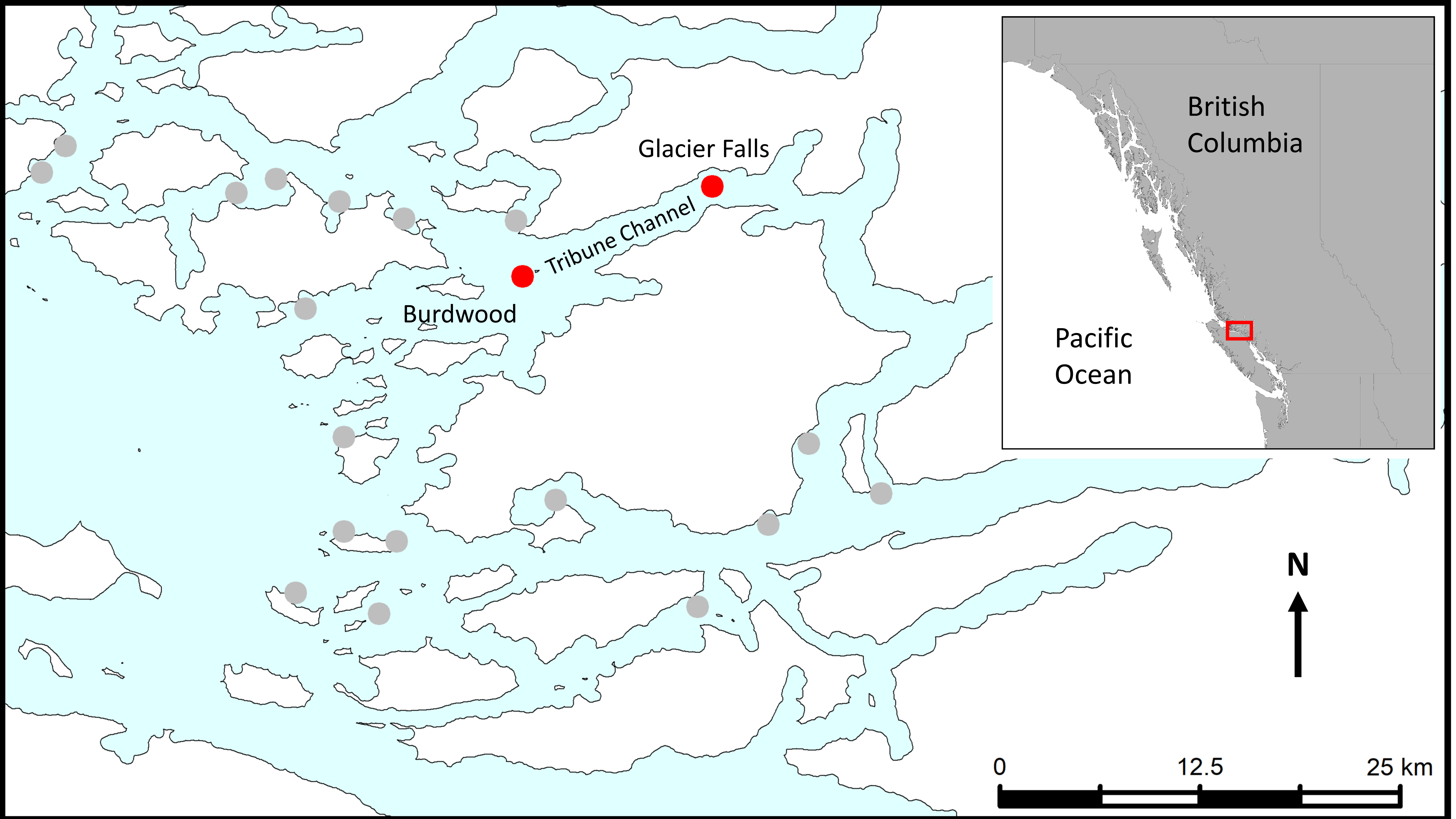}}

\subfloat[]{
\begin{tikzpicture}
\draw (0,0) rectangle (11,1);
\node (glacier) [draw=black, fill=gray!40!white] at (3, 0.25) {Glacier Falls};
\node (burdwood) [draw=black, fill=gray!40!white] at (8, 0.25) {Burdwood};
\draw[->, ultra thick,>=stealth] (4.5,0.75) -- (6.5,0.75);
\node (0) [draw=none, fill=none] at (7,-0.25) {$0$};
\node (l) [draw=none, fill=none] at (9,-0.25) {$L$};
\node (x0) [draw=none, fill=none] at (3, -0.25) {$x_0$};
\node (v) [draw=none, fill=none] at (5.5, 0.5) {$v$};
\end{tikzpicture}}
\caption{The two salmon farms that are used to calculate the time and probability of arrival for sea lice dispersing between farms. a) A map of the Broughton Archipelago with all active farms from 2009 shown in grey, and the two farms used in this study highlighted in red. The release farm is the eastern farm, Glacier Falls, which is located in Tribune Channel and the receiving farm is the western farm, Burdwood. b) The one dimensional representation of Tribune Channel used in the mathematical analysis. Note that the position of the farms has been  switched so the advective coefficient, $v$, is positive.   \label{fig:maps}}
\end{figure}
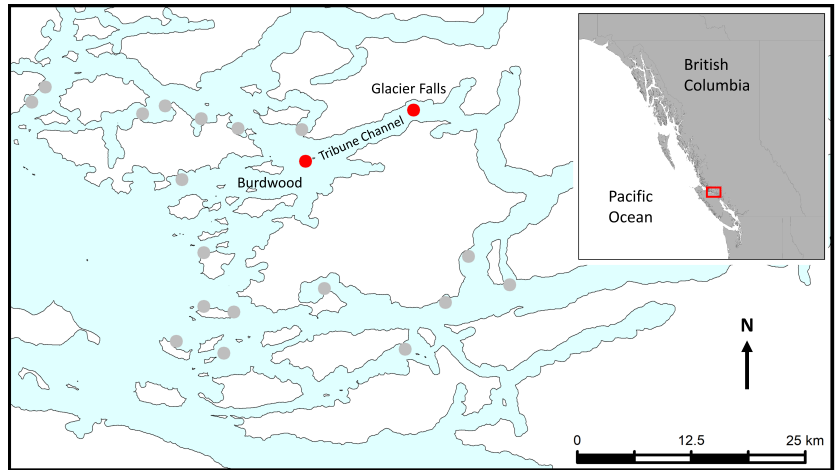
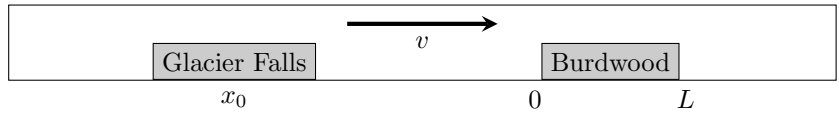

In dense salmon farming regions such as Norway, Scotland, and Canada there is evidence that sea lice populations on salmon farms are connected via larval dispersal and thus act as a connected metapopulation \citep{Adams2015, Aldrin2013, Cantrell2018, Cantrell2021, Aldrin2017}. In Norway, where most salmon farms are located in fjords along the coast, seaway distance has been used as a simple measure of farm connectivity over a large scale \citep{Aldrin2013}. At a smaller scale, hydrodynamic models have been used to measure the level of connectivity between farms \citep{Foreman2015, Cantrell2018, Adams2015, Samsing2017} in several salmon farming regions. Hydrodynamic models simulate ocean currents and can then be coupled with particle tracking models to determine how sea lice disperse when released from a farm \citep{Stucchi2011}.

The degree of interfarm connectivity, even between two farms, can have large consequences in terms of treating for sea louse outbreaks, and can even lead to chaotic dynamics under threshold treatment regimes \citep{Peacock2016}. Calculating the probability of sea lice dispersing to other farms is integral in determining which farms may be the largest sources of sea lice spread in a salmon farming region and thus which may be driving spread \citep{Cantrell2018, Harrington2020}. Lastly and perhaps most importantly, determining the probability of sea lice arrival onto other farms is critical in understanding where to place farms in a salmon farming region, or which to first remove.

To date, most of the research into the degree of connectivity between salmon farms has either been region specific, using hydrodynamical models \citep{Cantrell2018, Samsing2017} or at a large scale, with statistical analyses \citep{Aldrin2013, Kristoffersen2013}. Hydrodynamic models can be very useful in determining connectivity in the specific regions for which they are run, but results from these specific regions may be difficult to generalise to other regions. Conversely, statistical analyses are useful at determining the broad drivers of spread over large regions but often do not allow for detailed investigations into how certain parameter interactions affect the degree of connectivity between two farms. Thus there are still many general questions surrounding interfarm connectivity that require new approaches to investigate.

In this paper, we aim to answer the following questions surrounding the probability of sea lice dispersing between salmon farms:
\begin{enumerate}[(i)]
\item How does the degree of cross-infection, giving by arrival probability, depend on the spacing between farms?
\item Are there scenarios where an intermediate spacing between farms leads to the highest level of cross-infection?
\item Does the relationship between cross-infection and farm spacing change in advection dominated versus diffusion dominated systems?
\item How does the maturation time for nauplii to develop into infectious copepodites affect cross-infection?
\end{enumerate}

%but do not allow for detailed analysis of specific questions, such as how does distance change connectivity. Simmilarly hydrodynamic models are very useful in answering squestions in the specific region that they are run, but cannot easily be generalised to other regions. Here we focus on developing a mechanistic model of sea lice spread between salmon farms to answer the following general questions:
%
%

In order to answer these questions it is necessary to have a mechanistic model that is both sufficiently simple to investigate analytically and numerically in detail, but sufficiently realistic to capture the essential components of ocean circulation and sea louse biology. For credible analysis the simple model must be fit to realistic sea lice dispersal data to ensure accurate estimates of oceanic advection and diffusion as well as sea louse maturation times. The approach here is to use numerical flows from a three dimensional computational hydrodynamic model, the Finite Volume Community Ocean Model (FVCOM), along with a particle tracking model, to fit a simple one dimensional analytical model that describes the movement of sea lice between two salmon farms in a channel in the Broughton Archipelago, British Columbia.

The paper is structured as follows. First, we develop a simple mechanistic model for the arrival time distribution of sea lice dispersing between two different salmon farms. The arrival time distribution is necessary to calculate the level of cross-infection between farms, which is given by the probability of arrival for sea lice dispersing between salmon farms. We begin by presenting the analytical results for simple particles dispersing, ignoring the maturation required for sea lice to become infectious, before presenting the full arrival time distribution for sea lice that encompasses the non-infectious nauplius stage and infectious copepodite stage. Next, we calculate the arrival time directly using the numerical flows from FVCOM coupled with a particle tracking model to fit our simple mechanistic model and find parameter estimates. Lastly, we use our parameterized mechanistic model to investigate the questions (i)-(iv) surrounding cross-infection and farm placement.

\section{Methods}

\subsection{Study Area}

The Broughton Archipelago is a group of islands located between the northeast coast of Vancouver Island and the mainland of British Columbia (Figure \ref{fig:maps}). The Broughton Archipelago has several active salmon farms and has been at the center of the debate of the effect of sea lice on wild salmon \citep{Krkosek2005, Krkosek2006, Brooks2005, Krkosek2006a, Marty2010, Krkosek2007, Krkosek2011, Brooks2006, Krkosek2008, Riddell2008}. The rivers in the Broughton are major migration routes for both pink and chum salmon, and sea lice from salmon farms in this region have contributed to population level declines in pink salmon \citep{Krkosek2007}. Currently, certain salmon farms are being removed under a new agreement between the governments of British Columbia and the Kwikwasut'inuxw Haxwa'mis, 'Namgis, and Mamalilikulla First Nations \citep{Brownsey2018}. The abundance of sea lice data from counts on farmed and wild salmon as well as the complex hydrodynamical particle tracking simulations run for this region make the Broughton Archipelago an ideal area to investigate the cross-infection of sea lice between salmon farms.

\subsection{Mechanistic model}
\label{sec:model}

In order to calculate the arrival time of sea lice travelling between salmon farms and the probability of arrival, we first need a model for how sea lice disperse along a channel and arrive at a salmon farm. We begin by ignoring the maturation time required for newly released nauplii to develop into infectious copepodites to gain a comprehensive understanding of the arrival time distribution and to simplify the details of the initial mathematical analysis. Therefore we are assuming that all sea lice released from a salmon farm are infectious and model their dispersal using the following advection-diffusion equation:

\begin{align}
\frac{\partial}{\partial t}p(x,t) & =-\underbrace{\frac{\partial}{\partial x}\left(vp(x,t)\right)}_\text{advection}+\underbrace{\frac{\partial^{2}}{\partial x^{2}}\left(Dp(x,t)\right)}_\text{diffusion}-\underbrace{\mu p(x,t)}_\text{mortality}-\underbrace{h(x)\alpha p(x,t)}_\text{arrival onto farm} \label{eq:pde1}\\
p(x,0) & =p_0(x)\\
h(x) &= \begin{cases} 1 & x\in[0,L]\\
0 & \text{otherwise}
\end{cases},
\end{align}
where $p(x,t)$ is the density of sea lice at position $x$ and time $t$, $p_0(x)$ is the initial distribution of sea lice, and the farm at which sea lice are arriving is located between $[0,L]$. The rate at which sea lice arrive onto the farm, through successful attachment to salmonid hosts, is given by $\alpha$; the mortality rate of lice is $\mu$; advection, which represents the general seaward flow due to river output, is given by $v$; and diffusion, representing mixing due to winds and tidal flow, is given by $D$. This equation has previously been used to model sea lice movement in the Broughton Archipalego, to demonstrate the distribution of sea lice on wild salmon caused by salmon farms along salmon migration routes \citep{Krkosek2005, Krkosek2006, Peacock2020}. The necessary boundary conditions that accompany equation \ref{eq:pde1} are:

\begin{align}
\lim_{x\rightarrow-\infty} p(x,t)&=0\\
\lim_{x\rightarrow\infty} p(x,t)&=0\\
\lim_{x\rightarrow-\infty}\frac{\partial}{\partial x} p(x,t)&=0\\
\lim_{x\rightarrow\infty} \frac{\partial}{\partial x} p(x,t)&=0
\end{align}

To calculate the time of arrival onto the farm, we first rescale the density of lice by their probability of survival up to time $t$, $p(x,t)=e^{-\mu t}\tilde{p}(x,t)$, so that $\tilde{p}(x,t)$ represents the probability density that lice are in the channel at position $x$ at time $t$ given that they have survived, and $e^{-\mu t}$ is the probability that they have survived up to time $t$. The reasoning behind this rescaling is that now when we are tracking $\tilde{p}(x,t)$, the probability density function for the movement of lice that have survived, the only way that lice can be removed from the channel is by arriving onto the farm. The equation describing $\tilde{p}(x,t)$ is

\begin{align}
\frac{\partial}{\partial t}\tilde{p}(x,t) & =-\frac{\partial}{\partial x}\left(v\tilde{p}(x,t)\right)+\frac{\partial^{2}}{\partial x^{2}}\left(D\tilde{p}(x,t)\right)-h(x)\alpha \tilde{p}(x,t) \label{eq:tildep}\\
\tilde{p}(x,0) & =p_0(x)\\
h(x) &= \begin{cases} 1 & x\in[0,L]\\
0 & \text{otherwise}
\end{cases},
\end{align}
with the same necessary boundary conditions as before.

Let $T$ be the random variable describing the time of arrival onto the farm. We are interested in calculating $f(t)$, the distribution of arrival times, where $\int_0^tf(\tau)d\tau=\Pr(T<t)$. Now consider $\int_{-\infty}^{\infty} \tilde{p}(x,t)dx$. This is the probability  that lice are still in the water column and have not yet arrived onto the farm, thus $\int_{-\infty}^{\infty} \tilde{p}(x,t)dx=\Pr(T>t)=1-\Pr(T<t)$. The arrival time distribution $f(t)$ can therefore be given by

\[f(t)=-\frac{d}{dt}\int_{-\infty}^{\infty} \tilde{p}(x,t)dx.\]

If we integrate equation \ref{eq:tildep} in space from $-\infty$ to $\infty$, then the advection and diffusion terms disappear due to the boundary conditions and we are left with 

\[\frac{d}{dt}\int_{-\infty}^{\infty} \tilde{p}(x,t)dx=-\alpha\int_0^L \tilde{p}(x,t) dx.\]
Substituting this equation into the one for $f(t)$ we find that 

\begin{equation}
f(t)=\alpha \int_0^L \tilde{p}(x,t)dx.
\end{equation}

Therefore in order to solve $f(t)$ we must in turn solve $\tilde{p}(x,t)$. Before turning our attention to this solution, there are a couple details which are important to note. First, because $f(t)$ is the distribution of arrival times of sea lice that arrive on the farm, $\int_0^{\infty}f(t)dt$ will only equal 1 if all lice are eventually arrive onto the farm. While this will be the case if $v=0$, if $v>0$ or $v<0$ then this need not be the case. In fact for sea lice passing by salmon farms, the arrival rate $\alpha$ will probably be quite small, as farms are often located on the edge of large channels, and we are approximating the entire channel with a one dimensional domain. Therefore most of the lice released will not arrive onto the farm, an assumption which we will make explicit in the following section.  Second, because we have removed mortality from the equation describing $\tilde{p}(x,t)$, $f(t)$ is the probability density of arrival at time $t$, given that lice survive up to time $t$. The density of lice that survive up to time $t$ and then arrive onto the farm is $e^{-\mu t} f(t)$. 

\subsubsection{Calculating arrival time via asymptotic analysis}
\label{sec:asymp}

The solution, $\tilde{p}(x,t)$, to equation \ref{eq:tildep} is difficult to solve exactly and so to find an analytical solution to $\tilde{p}(x,t)$ and $f(t)$ we perform an asymptotic analysis and solve the first order solution. For simplicity of asymptotic analysis we will assume that $p_0(x)=\delta(x-x_0)$, so that all lice are initially released from another farm at position $x_0$. To find a small parameter around which to perform the asymptotic analysis we first need to non-dimensionalize our system. There are many different possibilities for non-dimensionalization, but in our case we choose to non-dimensionalize time as $\tilde{t}=\frac{D}{L^2}t$ and space as $\tilde{x}=\frac{x}{L}$. Using these non-dimensional parameters we can re-write equation \ref{eq:tildep} as

%This assumption ignores the maturation time required for lice to become infective, but the assumption will be relaxed in a later section, and the main difference in the asymptotic analysis will be the use of a Green's function, over which the initial distribution will be integrated. 

\begin{align}
\frac{\partial}{\partial \tilde{t}}\tilde{p}(\tilde{x},\tilde{t}) & =-\frac{\partial}{\partial \tilde{x}}\left(\frac{vL}{D}  \tilde{p}(\tilde{x},\tilde{t})\right)+\frac{\partial^{2}}{\partial \tilde{x}^{2}}\tilde{p}(\tilde{x},\tilde{t})-h(\tilde{x})\frac{\alpha L^2}{D} \tilde{p}(\tilde{x},\tilde{t}) \label{eq:nondim}\\
\tilde{p}(\tilde{x},0) & =\delta(\tilde{x}-\frac{x_0}{L})\\
h(\tilde{x}) &= \begin{cases} 1 & \tilde{x}\in[0,1]\\
0 & \text{otherwise}
\end{cases}.
\end{align}

Along with non-dimensionalizing the equations for $\tilde{p}(\tilde{x},\tilde{t})$, we must also write $f(t)$ in terms of its non-dimensional form so that it is clear how to redimensionalize $f(t)$ to fit to data. Previously, we demonstrated that $f(t)$ could be calculated as 

\[f(t)=-\frac{d}{dt}\int_{-\infty}^{\infty} \tilde{p}(x,t) dx.\]

In terms of the new non-dimensional time and space variables, $\tilde{t}$ and $\tilde{x}$, this can be rewritten as 

\[f(\tilde{t})=-\frac{D}{L}\frac{d}{d\tilde{t}} \int_{-\infty}^{\infty} \tilde{p}(\tilde{x},\tilde{t})d\tilde{x}.\]

From the formulation of $\tilde{p}(\tilde{x},\tilde{t})$ in equation (\ref{eq:nondim}) we can see that the non-dimensional rate of removal of $\tilde{p}$ will be $-d/d\tilde{t}\int_{-\infty}^{\infty} \tilde{p}(\tilde{x},\tilde{t})d\tilde{x}$, in the same manner as the dimensional removal rate, $f(t)$, was calculated in the previous section. Therefore if we let \[\tilde{f}(\tilde{t})=-\frac{d}{d\tilde{t}} \int_{-\infty}^{\infty} \tilde{p}(\tilde{x},\tilde{t})d\tilde{x}\] be the non-dimensional arrival time distribution, then we can write the dimensional arrival time as

\begin{equation}
f(\tilde{t})=\frac{D}{L}\tilde{f}(\tilde{t}).
\end{equation}

In the non-dimensionalization process of $\tilde{p}(\tilde{x},\tilde{t})$ three dimensionless parameters appear: $x_0/L$, $vL/D$,  and $\alpha L^2/D$. In the Broughton Archipelago, advection and diffusion have previously been estimated as $v=0.0645$ km/hr and $D=0.945$ $\text{km}^2/\text{hr}$ (Table \ref{table:paramestcomp}), and the length of the average farm is around $L=0.1$ km.
To roughly estimate the magnitude of the arrival rate in 1 dimension, $\alpha$, we must first make some assumptions about the arrival rate of sea lice over a salmon farm in two dimensions. Let $\beta$ be the actual rate of arrival of lice onto a farm when they are in the water column directly over a farm. A recent lab based experimental study found that 12-56\% of copepodites can attach to salmon after one hour in semi stagnant water \citep{Skern-Mauritzen2020}. Using a simple exponential waiting time model, $P(t)=1-e^{-\beta t}$, where $P$ is the probability of attachment and $\beta$ is the attachment rate, an upper estimate of the attachment rate would be $\beta=0.821/\text{hr}$. To calculate $\alpha$, we assume that the ratio of $\alpha/\beta=0.012$, which in physical terms means that the two dimensional area taken up by the farm is roughly 0.012 of the area of the channel at the location of the farm. At the particular farm we fit the model to the farm is roughly 50m wide and the channel is 4km wide, and 0.05/4=0.0125). When we fit the model we find that the estimate of $\alpha/\beta$ in fact ranges from 0.012 to 0.006 (Table \ref{table:paramest}). Thus taking the values of $\alpha/\beta=0.012$ and $\beta=0.821$ as rough maximum estimates, we assume $\alpha \leq 0.01$. Based on these parameter estimates and assumptions we choose $\alpha L^2/D$ to be the small parameter around which we perform our asymptotic approximation.  

Let $z=x_0/L$, $\epsilon=\alpha L^2/D$ ($<1.1\times10^{-4}$) and $\omega=vL/D$ ($6.83\times 10^{-3}$). Then we can rewrite equation \ref{eq:nondim} as 

\begin{align}
\frac{\partial}{\partial \tilde{t}}\tilde{p}(\tilde{x},\tilde{t}) & =-\frac{\partial}{\partial \tilde{x}}\left(\omega  \tilde{p}(\tilde{x},\tilde{t})\right)+\frac{\partial^{2}}{\partial \tilde{x}^{2}}\tilde{p}(\tilde{x},\tilde{t})-\epsilon h(\tilde{x}) \tilde{p}(\tilde{x},\tilde{t}) \label{eq:nondim2}\\
\tilde{p}(\tilde{x},0) & =\delta(\tilde{x}-z)\\
h(\tilde{x}) &= \begin{cases} 1 & \tilde{x}\in[0,1]\\
0 & \text{otherwise}
\end{cases}.
\end{align}

In terms of $\tilde{f}(\tilde{t})$, if we integrate both sides of equation \ref{eq:nondim2} on space from $-\infty$ to $\infty$, then we can write \[\tilde{f}(\tilde{t})=\epsilon \int_0^1 \tilde{p}(\tilde{x}, \tilde{t})d\tilde{x}.\]

We assume that $\tilde{p}(\tilde{x},\tilde{t})$ can be expressed as a regular asymptotic expansion in epsilon, \[\tilde{p}(\tilde{x},\tilde{t})=\tilde{p}_0(\tilde{x},\tilde{t})+\epsilon \tilde{p}_1(\tilde{x},\tilde{t}) +O(\epsilon^2)\]
and then can express $\tilde{f}(\tilde{t})$ as  \[\tilde{f}(\tilde{t})=\epsilon \int_0^1 \tilde{p}_0(\tilde{x},\tilde{t})d\tilde{x}+O(\epsilon^2).\]

Substituting the expansion for $\tilde{p}(\tilde{x},\tilde{t})$ into equation \ref{eq:nondim2} and matching terms of order $\epsilon^0$, we have 

\begin{align}
\frac{\partial}{\partial \tilde{t}}\tilde{p}_0(\tilde{x},\tilde{t}) & =-\frac{\partial}{\partial \tilde{x}}\left(\omega  \tilde{p}_0(\tilde{x},\tilde{t})\right)+\frac{\partial^{2}}{\partial \tilde{x}^{2}}\tilde{p}_0(\tilde{x},\tilde{t})\label{eq:asym0}\\
\tilde{p}_0(\tilde{x},0) & =\delta(\tilde{x}-z),
\end{align}
which has the solution 
\begin{equation}
\tilde{p}_0(\tilde{x},\tilde{t})=\frac{1}{\sqrt{4 \pi \tilde{t}}}e^{-(\tilde{x}-z-\omega \tilde{t})^2/4\tilde{t}}.\label{eq:firstorder}
\end{equation}

%Corresponding to the zero-th order approximation in $\tilde{p}(x,t)$, we can look at the zero-th order approximation of the arrival time distribution, $f(t)$. Recall that $f(t)=-d/dt \int_{-\infty}^{\infty}\tilde{p}(x,t)dx$, and so we can use the same $\epsilon$ expansion for $f(t)$, where 
%\begin{align}
%f(t)&=-\frac{d}{dt} \int_{-\infty}^{\infty}\tilde{p}_0(x,t)+\epsilon \tilde{p}_1(x,t) +O(\epsilon^2) dx\\
%&=\underbrace{-\frac{d}{dt} \int_{-\infty}^{\infty}\tilde{p}_0(x,t)dx}_\text{$f_0(t)$} +\epsilon\underbrace{\left(-\frac{d}{dt} \int_{-\infty}^{\infty}\tilde{p}_1(x,t)dx\right)}_\text{$f_1(t)$}+O(\epsilon^2)\label{eq:asymf}
%\end{align}
%Integrating $\tilde{p}_0(x,t)$ across space we find that $f_0(t)=-d/dt \int_{-\infty}^{\infty}\tilde{p}_0(x,t)dx=0$. What we can see here is that if the rate of arrival onto the farm is 0, then the probability of lice arriving onto the farm is also 0.
%
%Moving on to the first order expansion of $\tilde{p}(x,t)$ in $\epsilon$, after matching terms of order $\epsilon^1$ we have
%
%\begin{align}
%\frac{\partial}{\partial t}\tilde{p}_1(x,t) & =-\frac{\partial}{\partial x}\left(\omega  p_1(x,t)\right)+\frac{\partial^{2}}{\partial x^{2}}p_1(x,t)-h(x)p_0(x,t)\label{eq:asym1}\\
%p_1(x,0) & =0\\
%h(x) &= \begin{cases} 1 & x\in[0,1]\\
%0 & \text{otherwise}
%\end{cases}.
%\end{align}
%
%Now instead of solving for $p_1(x,t)$ explicitly, we move directly to solving the first order approximation of $f(t)$, $f_1(t)$. We solve $f_1(t)$ via 

Therefore $\tilde{f}(\tilde{t})$, up to order $\epsilon^2$, is given by

\begin{align}
\tilde{f}(\tilde{t})&=\epsilon\int_0^1p_0(\tilde{x},\tilde{t})d\tilde{x}+O(\epsilon^2) \\
&=\epsilon\int_0^1\frac{1}{\sqrt{4 \pi \tilde{t}}}e^{-(\tilde{x}-z-\omega \tilde{t})^2/4\tilde{t}} d\tilde{x}+O(\epsilon^2)\label{eq:f1}
\end{align}

Returning to our original dimensional parameters, the dimensional form of the arrival time distribution is

\begin{align}
f(t)&=\alpha\int_0^L\frac{1}{\sqrt{4 \pi D t}}e^{-(x-x0-vt)^2/4Dt} dx\\
&=\frac{\alpha}{2}\left(\erf\left(\frac{x_0+vt}{\sqrt{4Dt}}\right) -\erf\left(\frac{x_0+vt-L}{\sqrt{4Dt}}\right)\right).
\end{align}

\subsubsection{Including survival and maturation}

In the previous section we ignored the fact that sea lice larvae can be divided into two main stages: a non-infectious nauplius stage, and an infectious copepodite stage. In the nauplius stage, sea lice larvae cannot attach to salmonid hosts even if they come in close contact, it is only in the copepodite stage that sea lice are able to attach to hosts. To capture this infectious stage, we model the densities of the nauplius ($p_n(x,t)$) and copepodite ($p_c(x,t)$) stages with the following differential equations:

\begin{align}
\frac{\partial}{\partial t}p_n(x,t) & =-\underbrace{\frac{\partial}{\partial x}\left(vp_n(x,t)\right)}_\text{advection}+\underbrace{\frac{\partial^{2}}{\partial x^{2}}\left(Dp_n(x,t)\right)}_\text{diffusion}-\underbrace{\mu_n p_n(x,t)}_\text{nauplius mortality} -\underbrace{m(t)p_n(x,t)}_\text{maturation}  \label{eq:pden}\\
p_n(x,0) & =\delta(x-x_0)\\
\frac{\partial}{\partial t}p_c(x,t) & =\underbrace{m(t)p_n(x,t)}_\text{maturation}  -\underbrace{\frac{\partial}{\partial x}\left(vp_c(x,t)\right)}_\text{advection}+\underbrace{\frac{\partial^{2}}{\partial x^{2}}\left(Dp_c(x,t)\right)}_\text{diffusion}\\&-\underbrace{\mu_c p_c(x,t)}_\text{copepodite mortality}-\underbrace{\alpha h(x) p_c(x,t)}_\text{arrival onto farm}\nonumber\\
p_c(x,0)&=0\\
h(x) &= \begin{cases} 1 & x\in[0,L]\\
0 & \text{otherwise}
\end{cases},\label{eq:pden2}
\end{align}
where $\mu_i$ is the mortality rate in stage $i$, and $m(t)$ is the maturation rate of nauplii to copepodites.

We are again interested in calculating the time it takes for larvae leaving one farm to arrive on another. Now that we have divided the larvae into a non-infectious stage and an infectious stage, the larvae must mature into the infectious stage in order to arrive onto the second farm. Therefore to calculate the arrival time of larvae onto the second farm, we are really interested in the removal rate of infectious larvae from the water column. However, we do not want to count infectious larvae that die, so first we need to separate out mortality from the two equations. Let $p_c(x,t)=e^{-\mu_ct}\tilde{p}_c(x,t)$ and $p_n(x,t)=e^{-\mu_n t}\tilde{p}_n(x,t)$, where $e^{-\mu_ct}$ and $e^{-\mu_nt}$ are the probabilities that lice survive up to time $t$ in the copepodite and nauplius stages respectively. In the previous section where there was only one stage the arrival time distribution was given by $f(t)=-d/dt\int_{-\infty}^{\infty}\tilde{p}(x,t)dx$, but now that there are two stages with different death rates the arrival time distribution will be given by

\begin{equation}
e^{-\mu_ct}f(t)=-e^{\mu_c t} \frac{d}{dt}\int_{-\infty}^{\infty}\tilde{p}_c(x,t)dx-e^{-\mu_n t}\frac{d}{dt}\int_{-\infty}^{\infty}\tilde{p}_n(x,t)dx.\label{eq:fn}
\end{equation}

To calculate the arrival time we can rewrite equations (\ref{eq:pden})-(\ref{eq:pden2}) as
\begin{align}
e^{-\mu_nt}\frac{\partial}{\partial t}\tilde{p}_n(x,t) & =-e^{-\mu_nt}\frac{\partial}{\partial x}\left(v\tilde{p}_n(x,t)\right)+e^{-\mu_nt}\frac{\partial^{2}}{\partial x^{2}}\left(D\tilde{p}_n(x,t)\right) -e^{-\mu_nt}m(t)\tilde{p}_n(x,t)\label{eq:tpden}\\
\tilde{p}_n(x,0) & =\delta(x-x_0)\\
e^{-\mu_ct}\frac{\partial}{\partial t}\tilde{p}_c(x,t) & =e^{-\mu_nt}m(t)\tilde{p}_n(x,t)  - e^{-\mu_ct}\frac{\partial}{\partial x}\left(v\tilde{p}_c(x,t)\right)+e^{-\mu_ct}\frac{\partial^{2}}{\partial x^{2}}\left(D\tilde{p}_c(x,t)\right)\nonumber\\
&-e^{-\mu_ct}\alpha h(x) \tilde{p}_c(x,t)\label{eq:tpc}\\
\tilde{p}_c(x,0)&=0\\
h(x) &= \begin{cases} 1 & x\in[0,L]\\
0 & \text{otherwise}
\end{cases}.\label{eq:tpden2}
\end{align}
Then adding them together, integrating over all space, and substituting the resulting expression into equation \ref{eq:fn} we have

%By integrating the above equations over the spatial domain, and adding the nauplius and copepodid equations together, we can track the rate of lice leaving the channel domain and arriving onto a farm. We add the equations together so that the maturation rate of nauplii to copepodids is ignored, and we have already accounted for removal due to death by rescaling each equation by its survival function. When the two equations are added and integrated over all space we are left with 
%
%\begin{equation}
%e^{-\mu_n t}\frac{d}{dt}\int_{-\infty}^{\infty}\tilde{p}_n(x,t)dx+e^{-\mu_c t}\frac{d}{dt}\int_{-\infty}^{\infty}\tilde{p}_c(x,t)dx=-e^{-\mu_ct}\alpha\int_0^L\tilde{p}_c(x,t)dx.
%\end{equation}
%
%The exponentials in front of each term are the probabilities that the larvae survive up to time $t$, and so the rate that larvae are removed from the system and arriving on the second farm, given by the arrival time density function, is

\[f_c(t)=\alpha\int_0^L\tilde{p}_c(x,t)dx,\] where we use the subscript $c$ to denote the arrival time of lice which have matured into copepodites.

Once again, to calculate the arrival time density we will need to solve the equations governing the lice distribution in the channel using an asymptotic analysis, this time with the addition of Green's functions.

First, let us formulate the copepodid density, $\tilde{p}_c(x,t)$, in terms of a Green's function. The Green's function describes the movement of copepodites, as described by equation (\ref{eq:tpc}), but without the source of maturing nauplii. The Green's function is then convolved with the source function: the maturing nauplii which are entering the copepodid stage. The copepodid density  can then be written as
\begin{equation}\tilde{p}_c(x,t)=\int_0^t \int_{-\infty}^{\infty} G(x-\xi,t-\tau)s(\xi,\tau)d\xi d\tau,\label{eq:pc}\end{equation} where $s(\xi,\tau)=e^{(\mu_c-\mu_n)\tau} m(\tau)\tilde{p}_n(\xi,\tau)$ and $G(x,t)$ solves 

\begin{align}
\frac{\partial}{\partial t}G(x,t) & =-\frac{\partial}{\partial x}\left(vG(x,t)\right)+\frac{\partial^{2}}{\partial x^{2}}\left(DG(x,t)\right)-\alpha h(x) G(x,t)\label{eq:green}\\
G(x,0)&=\delta(x)\\
h(x) &= \begin{cases} 1 & x\in[0,L]\\
0 & \text{otherwise}
\end{cases}.\label{eq:green2}
\end{align}

Similar to before, the equation governing $G(x,t)$ is difficult to solve directly, and so we non-dimensionalize the equations and then perform an asymptotic analysis in a small parameter. We non-dimensionalize equations (\ref{eq:green})-(\ref{eq:green2}) and non-dimensionalize the formula for $\tilde{p}_c(x,t)$ (equation (\ref{eq:pc})) directly. As before, let $\tilde{t}=\frac{D}{L^2}t$ and $\tilde{x}=\frac{x}{L}$, then the nauplius and copepodid system can be reformulated in a non-dimensional form as:

\begin{align}
\frac{\partial}{\partial \tilde{t}}\tilde{p}_n(\tilde{x},\tilde{t}) & =-\frac{\partial}{\partial \tilde{x}}\left(\frac{vL}{D} \tilde{p}_n(\tilde{x},{t})\right)+\frac{\partial^{2}}{\partial \tilde{x}^{2}}\tilde{p}_n(\tilde{x},\tilde{t}) -\frac{L^2}{D}m(\tilde{t})p_n(\tilde{x},\tilde{t})\\
\tilde{p}_n(\tilde{x},0) & =\delta(\tilde{x}-\frac{x_0}{L})\\
\frac{\partial}{\partial \tilde{t}}G(\tilde{x},\tilde{t}) & =-\frac{\partial}{\partial \tilde{x}}\left(\frac{vL}{D}G(\tilde{x},\tilde{t})\right)+\frac{\partial^{2}}{\partial \tilde{x}^{2}}G(\tilde{x},\tilde{t})-\frac{\alpha L^2}{D} h(x) G(x,t)\label{eq:greennondim}\\
G(\tilde{x},0)&=\delta(\tilde{x})\\
h(x) &= \begin{cases} 1 & x\in[0,L]\\
0 & \text{otherwise}
\end{cases}\\
\tilde{p}_c(\tilde{x},\tilde{t})&=\frac{L^3}{D}\int_0^{\tilde{t}} \int_{-\infty}^{\infty} G(\tilde{x}-\tilde{\xi},\tilde{t}-\tilde{\tau})e^{(\mu_c-\mu_n)(L^2\tilde{\tau}/D)} m(\tilde{t})\tilde{p}_n(\tilde{\xi},\tilde{\tau}) d\tilde{\xi} d\tilde{\tau}.
\end{align}

Similar to the previous section, we also need to write $f(t)$ in terms of the new non-dimensional space and time variables. Rescaling equation (\ref{eq:fn}) we have
\begin{equation}
e^{-\mu_c(L^2\tilde{t}/D)}f_c(\tilde{t})=\frac{D}{L}\left(-e^{\mu_c (L^2\tilde{t}/D)} \frac{d}{d\tilde{t}}\int_{-\infty}^{\infty}\tilde{p}_c(\tilde{x},\tilde{t})d\tilde{x}-e^{-\mu_n (L^2\tilde{t}/D)}\frac{d}{d\tilde{t}}\int_{-\infty}^{\infty}\tilde{p}_n(\tilde{x},\tilde{t})d\tilde{x}\right),
\end{equation}
so if we let the non-dimensional version of our arrival time distribution be given by
\begin{equation}
e^{-\mu_c(L^2\tilde{t}/D)}\tilde{f}_c(\tilde{t})=\left(-e^{\mu_c (L^2\tilde{t}/D)} \frac{d}{d\tilde{t}}\int_{-\infty}^{\infty}\tilde{p}_c(\tilde{x},\tilde{t})d\tilde{x}-e^{-\mu_n (L^2\tilde{t}/D)}\frac{d}{d\tilde{t}}\int_{-\infty}^{\infty}\tilde{p}_n(\tilde{x},\tilde{t})d\tilde{x}\right)\label{eq:nondimfn}
\end{equation}
then the relationship between the dimension and non-dimensional forms of the arrival time is
\[f_c(\tilde{t})=\frac{D}{L}\tilde{f}_c(\tilde{t}).\]

For our small parameter we again choose $\epsilon=\alpha L^2/D$, around which we perform our expansion, and $\omega=vL/D$ as the other non-dimensional parameter. We can then expand $G(\tilde{x},\tilde{t})=G_0(\tilde{x},\tilde{t})+\epsilon G_1(\tilde{x},\tilde{t})+O(\epsilon^2)$, along with the corresponding $\tilde{p}_c(\tilde{x},\tilde{t})=\tilde{p}_{c0}(\tilde{x},\tilde{t})+\epsilon \tilde{p}_{c1}(\tilde{x},\tilde{t})+O(\epsilon^2)$.

Then using these new parameters and adding together the terms on the right hand side of equation (\ref{eq:nondimfn}), the non-dimensional version of the arrival time distribution, $\tilde{f}(\tilde{t})$, is 
\begin{equation}
\tilde{f}_c(\tilde{t})=\epsilon \int_0^1 \tilde{p}_{c0}(\tilde{x},\tilde{t})d\tilde{x}+O(\epsilon^2).
\end{equation}
Therefore to find $\tilde{f}_c(\tilde{t})$ we need to calculate $G_0(\tilde{x},\tilde{t})$ and $\tilde{p}_0(\tilde{x},\tilde{t})$. Focusing first on $G(\tilde{x},\tilde{t})$ we can match terms of order $\epsilon^0$ in equation (\ref{eq:greennondim}) to arrive at 

\begin{align}
\frac{\partial}{\partial \tilde{t}}G_0(\tilde{x},\tilde{t}) & =-\frac{\partial}{\partial \tilde{x}}\left(\omega G_0(\tilde{x},\tilde{t})\right)+\frac{\partial^{2}}{\partial \tilde{x}^{2}}G_0(\tilde{x},\tilde{t})\\
G_0(\tilde{x},0)&=\delta(\tilde{x})
\end{align}
which has the solution 
 
\begin{equation}
G_0(\tilde{x},\tilde{t})=\frac{1}{\sqrt{4 \pi \tilde{t}}}e^{-(\tilde{x}-\omega \tilde{t})^2/4\tilde{t}}\label{eq:G}
\end{equation}
with the corresponding solution in $\tilde{p}_{c0}(\tilde{x},\tilde{t})$,

\begin{equation}
\tilde{p}_{c0}(\tilde{x},\tilde{t})=\frac{L^3}{D}\int_0^{\tilde{t}} \int_{-\infty}^{\infty} G_0(\tilde{x}-\tilde{\xi},\tilde{t}-\tilde{\tau})e^{(\mu_c-\mu_n)(L^2\tilde{\tau}/D)} m(\tilde{t})\tilde{p}_n(\tilde{\xi},\tilde{\tau}) d\tilde{\xi} d\tilde{\tau}. \label{eq:p0}
\end{equation}

Therefore the arrival time distribution, up to order $\epsilon^2$, is given by  $\tilde{f}_c(\tilde{t})=$ $\epsilon \int_0^1 \tilde{p}_{c0}(\tilde{x},\tilde{t})d\tilde{x} +O(\epsilon^2)$ along with equations (\ref{eq:G}) and (\ref{eq:p0}). Before writing out $\tilde{f}_c(\tilde{t})$ explicitly, we will first redimensionalize the arrival time distribution, as this is what will be fit to data. In its dimensional form with the original time and space variables we have:

\begin{align}
f_c(t)& \approx \alpha\int_0^L \int_0^t \int_{-\infty}^{\infty}G_0(x-\xi,t-\tau)e^{(\mu_c-\mu_n)\tau}m(t)\tilde{p}_n(\xi,\tau)d\xi d\tau\\
G_0(x,t)&=\frac{1}{\sqrt{4\pi Dt}}e^{-(x-vt)^2/4Dt}\\
\tilde{p}_n(x,t)&=\frac{1}{\sqrt{4\pi Dt}}e^{-(x-vt)^2/4Dt}e^{-\int_0^t m(u)du}
\end{align}

\subsection{Coupled biological-physical particle tracking simulation}
\label{sec:hydrodeets}

To measure the arrival time for sea lice dispersing between farms in the Broughton Archipelago we use a bio-physical particle tracking simulation. This particle tracking simulation uses an underlying ocean circulation model, the Finite Volume Community Ocean Model (FVCOM) \citep{Chen2006}. The simulation period of the FVCOM was between March 1st and July 31st 2009, to coincide with the outmigration of juvenile pink and chum salmon in that year. More details on the FVCOM simulation can be found in \citet{Foreman2009} and \citet{Cantrell2018}, but briefly the FVCOM uses data on tides, wind, surface heating, and river discharge from the six major rivers in the Broughton as input to simulate three-dimensional ocean velocity, temperature and salinity. The FVCOM uses an unstructured grid to solve the necessary hydrodynamic equations, which allows for a more realistic simulation of ocean circulation near the complex coastlines of the Broughton Archipelago. The FVCOM currents arising from this 2009 simulation were compared with observations from twelve current meter moorings and found to be in relatively close agreement \citep{Foreman2009}.

Hourly output from the FVCOM model was used as input into an offline bio-physical particle tracking simulation, details of which can be found in \citet{Cantrell2018}. The physical component of the particle tracking simulation determines how sea lice particles move based on the current that they experience from the output of the FVCOM model, and the biological component determines how they survive and mature based on the local salinity and temperature that they experience. Particles are first released from farms as pre-infectious nauplii and then mature into infectious copepodites. The development time from nauplii to copepodite is based on the temperature ($T$) that particles experience, and is given by the simplified B\v{e}lehr\'{a}dek  function \begin{equation}\tau (T)=\left[\frac{\beta_1}{T-10+\beta_1\beta_2}\right]^2.\label{eq:devtime}
\end{equation} As a particle will experience different temperatures over its lifetime, to track maturity each particle is given a maturity value ($M$). The maturity value starts at 0 for a newly released nauplius and then updates via 
\begin{equation}
M_t=M_{t-1}+\Delta t/\tau (T). \label{eq:maturation}
\end{equation}
Once the maturity value, $M$, reaches 1, the particle molts into a copepodite. 

The survival probability of each particle is given by \[S(t)=e^{-\mu t},\] where the survival coefficient, $\mu$, is constant at 0.31 per day when salinity ($S$) is above 30 ppt for the nauplius stage, and at less than 30 ppt is given by \begin{equation}\mu = 5.11-0.16S.\label{eq:sal}
\end{equation} Once the particles mature into copepodites (with a maturity coefficient $\geq 1$), then the survival coefficient is constant at 0.22 per day. The constant survival in the mature copepodite stage is due to a lack of consensus among studies on how copepodite survival changes with temperature and salinity.

To determine the trajectories of sea lice originating from farms in the Broughton Archipelago, 50 particles were released per hour from each of the 20 active farms (during 2009) and tracked for 11 days using the offline particle tracking model. The first day of release was March 14, and the last day of release was July 20, 2009. In this paper, to fit the analytical arrival time distribution, we use the particles released on May 2, 2009 (CRD 50 in \citet{Cantrell2018}). We fit to data from this day as it was the one chosen as being representative of an “average” day by \citet{Cantrell2018}. However, we also present the results of a fit to a high connectivity day, May 11, 2009 (CRD 59 in \citet{Cantrell2018}) in Appendix \ref{sec:highconnectivity}. 

The 24 hours of particle releases (24 releases $\times$ 50 particles per release) on May 2, 2009 were combined into one cohort so that the time of release of the entire cohort begins at $t=0$. The amalgamation of 24 hours of releases on one day is to smooth out the effect that the tidal cycle may have on any given individual release. When fitting arrival time using the amalgamation of 24 hours of particle releases, the constant advection coefficient in the model captures  directional water movement due to river runoff and the diffusion in the model coefficient captures the average mixing due to tidal flow and wind currents. For this particular cohort, Kernel Density Estimates (KDEs) were created using the particle locations at every hour over the 11 days that they were tracked, for a total of 265 KDEs. The idea behind kernel density estimation is to create a distribution from individual particle locations by applying a smoothed Gaussian kernel around each particle location and then adding each kernel to create a distribution. The specific details behind the Kernel Density Estimation process for the sea lice particles can be found in \citet{Cantrell2018}.

\subsection{Model fitting}

In order to fit the analytical model of arrival time to the Kernel Density Estimates calculated from the particle tracking simulation we need to calculate a form of arrival time from the KDEs. The first step in this process is to determining which farm to set as the release farm for sea lice and which to use as the receiving farm. In the Broughton Archipelago the use of a one dimensional advection diffusion model to determine the distribution of sea lice on wild salmon from source farms has been fit to data mainly in Tribune channel and so to compare parameter estimates we choose farms also in Tribune. Our release farm is Glacier Falls, located in the center of Tribune channel, and our receiving farm is Burdwood, which lies at the opening of Tribune (see Figure \ref{fig:maps}).

The KDE represents the distribution of particles over space and has units of particles per kilometer squared. In order to convert this density distribution into a probability distribution, we first must divide the KDE by the total number of particles released, in this case 1200 (50 particles $\times$ 24 releases (1 per hour)). Then the new KDE represents the two dimensional probability distribution of particles, with the integral over the entire domain equal to one; this is now the two dimensional equivalent of $p(x,t)$. 

Recall that in the one dimensional mechanistic model, arrival time was calculated as \[f(t)=\alpha \int_0^L p(x,t)dx\] and so to calculate the arrival time for the particle tracking simulations we take the value of the rescaled KDE at the position of the receiving farm and multiply it by the area of the raster cell (approximately the size of the farm), which is $0.01\text{ km}^2$.  The only unknown quantity is the rate of arrival of lice onto the farm over which they are passing, which we call $\beta$, this is the two dimensional equivalent of $\alpha$. As detailed at the start of section \ref{sec:asymp}, we estimate that an upper estimate of the arrival rate is $\beta=0.821/\text{hr}$. We assume here that this rate is small enough such that number of lice that arrive onto the farm is small compared to the total number of lice in the rest of the domain, and thus we do not discount future KDEs by any proportion of lice that have potentially arrived on a farm. 
%
%A recent lab based experimental study found that 12-56\% of copepodites can attach to salmon after one hour based on temperature \citep{Skern-Mauritzen2020}. Using a simple exponential waiting time model, $P(t)=1-e^{-\beta t}$, where $P$ is the probability of attachment and $\beta$ is the attachment rate, an upper estimate of the attachment rate would be $\beta=0.821/\text{hr}$, but the attachment rate in field conditions could be quite different. No matter the rate, the key assumption that we make here is that the number of lice that arrive onto the farm is small compared to the total number of lice in the rest of the domain, and thus we do not discount future KDEs by any proportion of lice that have potentially arrived on a farm. 

The hydrodynamic equivalent of the arrival time distribution can then be calculated as \[f_h (t)=\beta \int_{farm}p_h(x,t)d\Omega,\] to which we want to fit our original arrival time distribution $f(t)=\alpha \int_0^L p(x,t)dx$. We could use our assumption of the arrival rate for the hydrodynamic model, $\beta$, to fit the arrival rate of the one dimensional model $\alpha$, however due to the uncertainties in $\beta$ we instead fit  \[\left(\frac{\alpha}{\beta}\right)\int_0^L p(x,t)dx\] to $\int_{farm}p_h (x,t)d\Omega,$ calculated from kernel density estimation. For simplicity we use $\beta=1$ rather than $\beta=0.821$ to calculate the arrival time in Figures 5.4 through 5.8.

%The $y$ axis of Figures 5.4, 5.5, 5.6, and 5.8 and the legend of Figure 5.7 in this paper are therefore scaled by  $(1/\beta)$.

%
%to kernel densities of lice above the farm, multiplied by the area of the farm, which is approximately $\int_{farm}p_h (x,t)d\Omega.$

There are three different mechanistic models that we fit to the arrival time from the Kernel Density Estimates: arrival time  of inert particles, arrival time of particles which have survived, and arrival time of sea lice particles that have survived and matured to their infective stage. These three models each use slightly different KDEs. For the inert particles, the KDEs are constructed solely from the positions of the sea lice particles at each time step. For the arrival time of only the particles that have survived up to $t$, the KDEs are constructed by weighting each particle by its individual survival coefficient, which depends on the local salinity that it has experienced, as described in the previous section. Details on this weighting can be found in \citet{Cantrell2018}. Lastly, for the arrival time of sea lice particles that have both survived up to time $t$ and matured into infectious copepodites, the KDEs are constructed using only particles that have a maturity value greater or equal to 0.8, and then these particles are again weighted by their survival coefficient during the construction of the KDEs. The value of 0.8 was chosen as it has been previously been found that using a maturity value of 1 may be overly sensitive to temperature \citep{Cantrell2018}.

To fit our arrival time models to the arrival time from the KDEs we use non-linear least squares, with the size of the farm and distance of release farm fixed at $L=0.1\text{ km}$ and $x_0=-13.5\text{ km}$ respectively. Maximum and minimum advection speeds were estimated using hourly snapshots of the KDEs and then used to bound the advection parameter $v$ during the non-linear least squares. The mortality rates ($\mu, \mu_n,\mu_c$) were constrained to lie within the maximum and minimum possible rates in the particle tracking model, and the ratio of the 1D to 2D arrival rates, $\alpha/\beta$ was constrained to be less than 0.0125, which is roughly the ratio of the width of the farm to the width of the channel. The best fit parameter estimates for all parameters can be found in Table \ref{table:paramest}. 

\begin{figure}
\centering
\includegraphics[width=9cm]{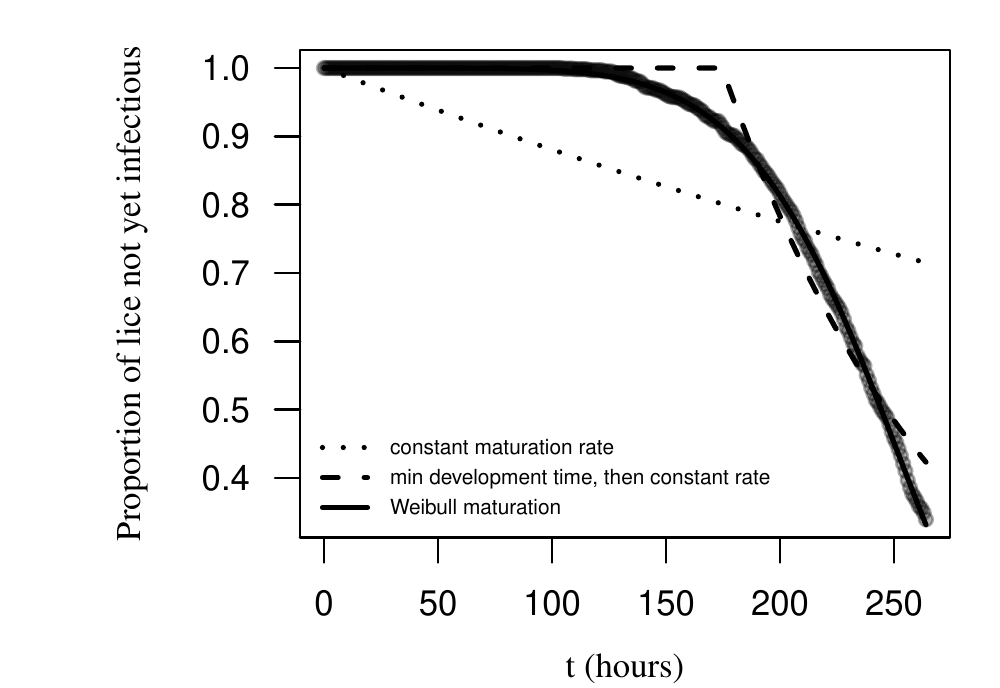}
\caption{The proportion of larvae that have not yet reached a maturation level of 0.8 in the hydrodynamic model, with the best fit lines for three different maturation functions. The dotted line is the maturation function corresponding to a constant maturation rate ($e^{-mt}$), the dashed line is the maturation function for a minimum development time followed by a constant maturation rate ($1-H(t-t_{\text{min}})(1-e^{-m(t-t_{\text{min})}}$), and the solid line is the Weibull maturation function ($e^{-\log(2)(t/\delta_m)^{\delta_s}}$).}
\label{fig:maturation}
\end{figure}

In the arrival time model which includes maturation, it is necessary to specify a maturation function in order to fit the model to the KDE data. We choose to use a Weibull maturation function, also used by \citet{Aldrin2017} to model sea lice maturation, as it gave a much better fit to our maturation data (Figure \ref{fig:maturation}) than the alternatives often used in the sea lice literature: a constant maturation rate \citep{Peacock2020, Krkosek2006} or strict minimum development time followed by a constant maturation rate \citep{Revie2005, Adams2015}. The Weibull distribution is a two parameter distribution and can be parameterized in a number of ways. We choose to follow \citet{Aldrin2017} and use the median time to development, $\delta_m$, and a shape parameter, $\delta_s$, to define the distribution. Using these parameters the maturation rate (often called the hazard rate in survival analysis) is \[m(t)=\log(2)\delta_s(\delta_m)^{-\delta_s}t^{\delta_s-1}.\]

\begin{sidewaystable}
\centering
\caption{Comparison of parameter estimates in this study to others in the Broughton Archipelago.}

\begin{tabular}{c l l l l l l}
\hline
Parameter & Description & Analytical Model & \citet{Krkosek2006} & \citet{Brooks2005} & \citet{Krkosek2005} & \citet{Foreman2009} \\
\hline
$v$ & Advection & 0.149 km/h & - &0.0648 km/h & 0.0379 km/h & 0.2088 km/h\\
$D$ & Diffusion & 0.617 $\text{km}^2$/h & 0.945 $\text{km}^2$/h & - & 0.321 $\text{km}^2$/h&-\\
$\mu_n$ & Nauplius mortality rate & 0.009/h & - & - & -&-\\
$\mu_c$ & Copepodite mortality rate &0.012/h & 0.0083/h & - & -&- \\
%$\delta_m$ & Median nauplius maturation time & 251h\\
\hline
\end{tabular}
\label{table:paramestcomp}
\end{sidewaystable}

\section{Results}

\subsection{Arrival time of inert particles}

First we present the arrival time distribution of inert sea lice particles, which do not have any survival or maturation characteristics associated with them, but are still confined to the top five meters of the water column. The formula for the arrival time distribution from the analytical model, given in section \ref{sec:model}, is 

\[f(t)=\alpha\int_0^L\frac{1}{\sqrt{4 \pi D t}}e^{-(x-x0-vt)^2/4Dt} dx.\]

\begin{figure}
\centering
\subfloat[]{
\includegraphics[width=7.5cm]{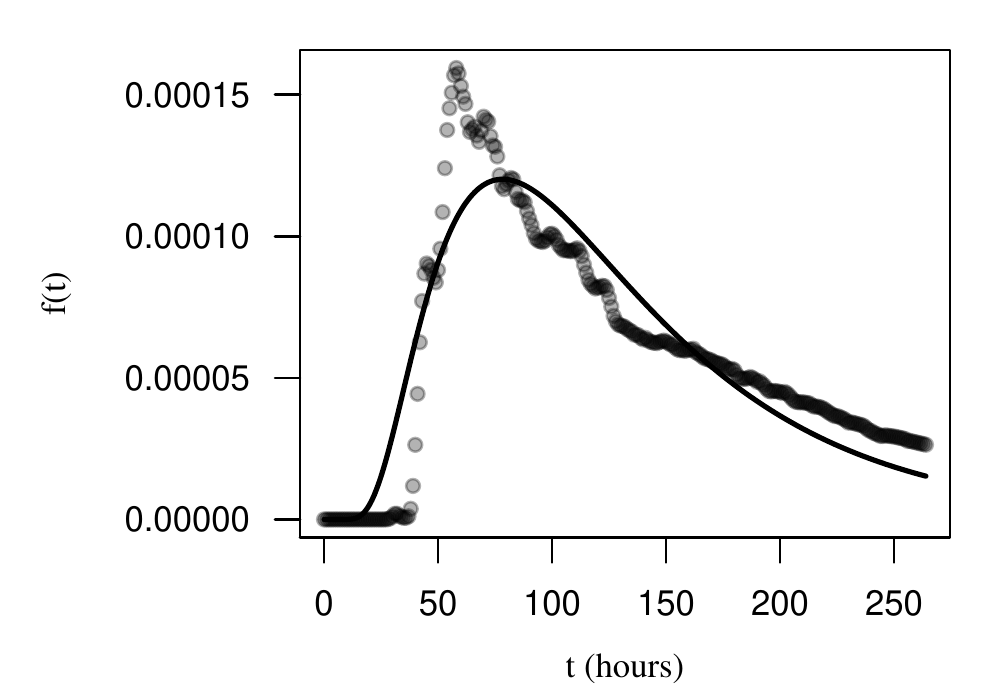}}
\subfloat[]{
\includegraphics[width=7.5cm]{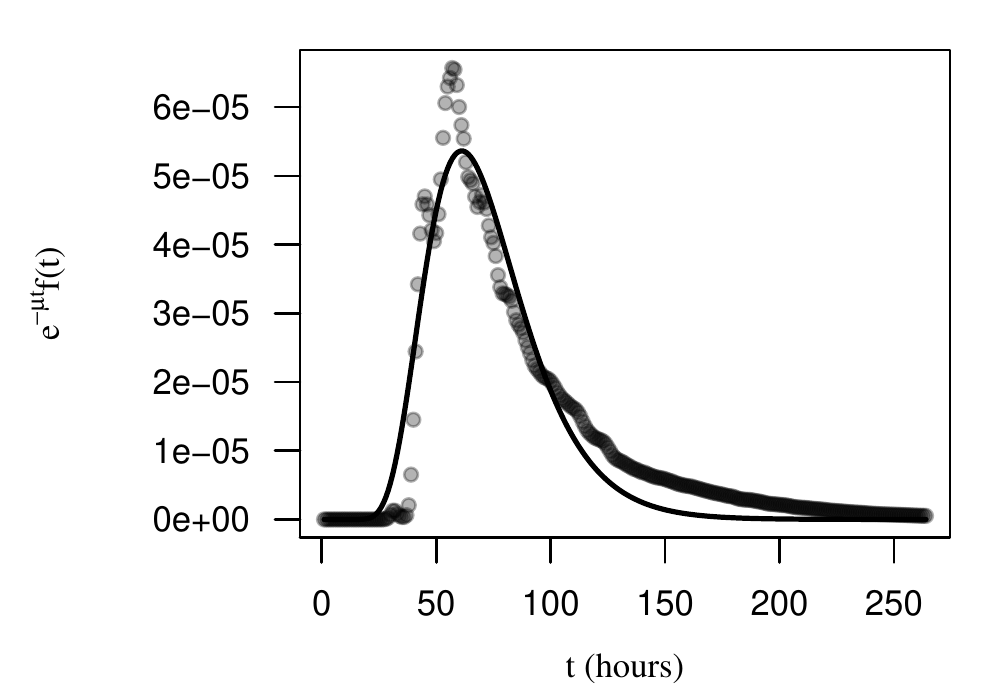}}

\subfloat[]{
\includegraphics[width=7.5cm]{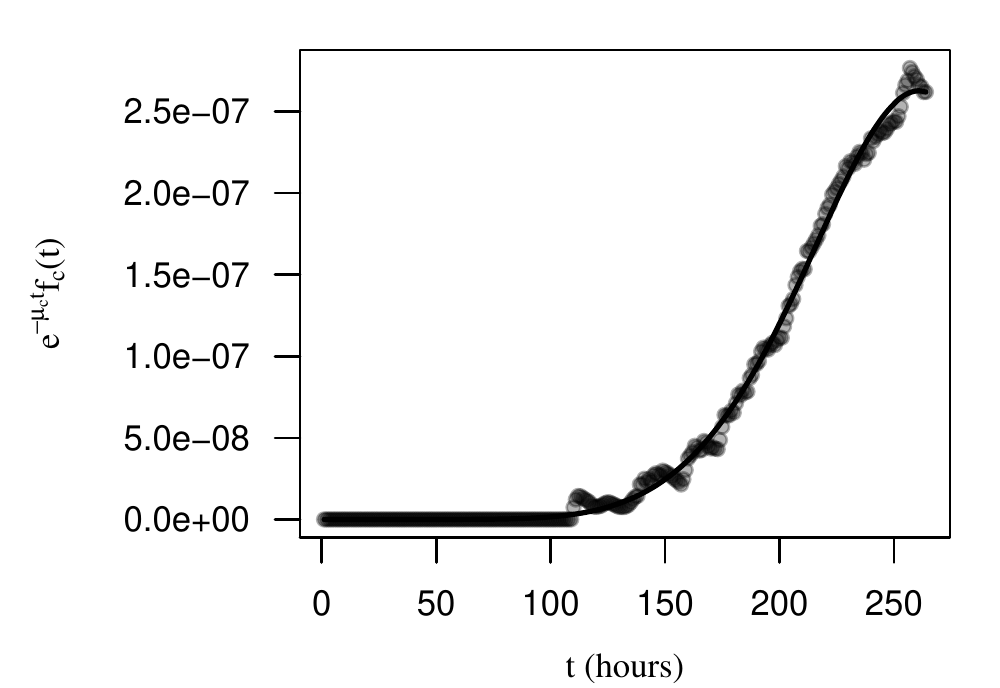}}

\caption{Arrival time distribution of a) inert particles, b) particles with survival included, and c) infectious copepodites. In a) the points are the arrival time densities calculated from the KDEs of the particle locations, in b) the points are the arrival time densities calculated from the KDEs of particle locations weighted by survival, and in c) the points are the arrival time densities calculated from infectious particle locations weighted by survival. The curves are the best fit lines of the corresponding analytical models fit to these points via non-linear least squares. Parameter estimates for the model are given in Table \ref{table:paramest}, with $\beta=1$.}
\label{fig:f}
\end{figure}

The fit of this distribution to the output from the hydrodynamic simulation can be seen in Figure \ref{fig:f}a), along with the best fit parameter estimates in Table \ref{table:paramest}.

\subsection{Arrival time including survival}

Next, we present the fit of the arrival time distribution of particles that have survived, to Kernel Density Estimates that are weighted by survival, as described in section \ref{sec:hydrodeets}. Therefore the distribution we are now fitting is \[e^{-\mu t} f(t)=\alpha e^{-\mu t} \int_0^L\frac{1}{\sqrt{4 \pi D t}}e^{-(x-x0-vt)^2/4Dt} dx.\]
The survival function, $e^{-\mu t}$, is the probability that a sea louse has survived up to time $t$, and $\int_{t-\epsilon}^{t+\epsilon} f(\tau)d\tau$ is the probability that a sea louse arrives on the second farm between $t-\epsilon$ and $t+\epsilon$, given that it has survived. The fit of $e^{-\mu t}f(t)$ to the hydrodynamic model is shown in Figure \ref{fig:f}b).  
 
%\begin{figure}
%\centering
%\includegraphics[width=9cm]{paperplot2.pdf}
%\caption{Arrival time distribution of particles with survival included. The points are the arrival time densities calculated from the hydrodynamic model with KDE estimates weighted by survival and the curve is the best fit line of the simple analytical model fit to these points via non-linear least squares. Parameter estimates for the model are given in Table \ref{table:paramest}, with $\beta=1$.}
%\label{fig:fs}
%\end{figure}

\subsection{Arrival time of sea lice (maturation and survival)}

Lastly, we present the fit of the arrival time distribution of infectious sea lice particles. In the hydrodynamic simulation these particles each mature and have a survival probability based on the local salinity and temperature that they experience over their lifetime. For the one dimensional mechanistic model, we are fitting the arrival time distribution of copepodites, $f(t)$, offset by the probability that a copepodite survives up to time $t$, $e^{-\mu_c t}$. In short, we are fitting

\begin{align}
e^{-\mu_c t} f_c(t)&=\alpha\int_0^L \int_0^t \int_{-\infty}^{\infty}G_0(x-\xi,t-\tau)e^{-\mu_c(t-\tau)} e^{-\mu_n \tau} m(\tau)p_n(\xi,\tau)d\xi d\tau\\
G_0(x,t)&=\frac{1}{\sqrt{4\pi Dt}}e^{-(x-vt)^2/4Dt}\\
p_n(x,t)&=\frac{1}{\sqrt{4\pi Dt}}e^{-(x-vt)^2/4Dt}e^{-\int_0^t m(u)du}\\
m(t)&=\log(2)\delta_s(\delta_m)^{-\delta_s}t^{\delta_s-1}
\end{align}
to the Kernel Density Estimates of sea lice particles that have survived and have matured from nauplii to infectious copepodites. The fit of $e^{-\mu_ct}f_c(t)$ is shown in Figure \ref{fig:f}c) and the best fit parameters are in Table \ref{table:paramest}. The probability of arrival of sea lice on the farm is given by \[\int_0^{\infty} e^{-\mu_c t}f_c(t)dt,\] which is how we will measure the level of cross infection between farms.

%\begin{figure}
%\centering
%\includegraphics[width=9cm]{paperplot3.pdf}
%\caption{Arrival time distribution of infectious copepodites. The points are the arrival time densities calculated from the hydrodynamic model and the curve is the best fit line of the simple analytical model fit to these points via non-linear least squares. Parameter estimates for the model are given in Table \ref{table:paramest}, with $\beta=1$.}
%\label{fig:fsm}
%\end{figure}

\begin{sidewaystable}
\centering
\caption{Parameter estimates of best fit models under non-linear least squares.}

\begin{tabular}{c l l l l}
\hline
Parameter & Description & Inert & With survival & With survival and maturation\\
\hline
$v$ & Advection & 0.143 km/h & 0.175 km/h & 0.149 km/h\\
$D$ & Diffusion & 0.371 $\text{km}^2$/h & 0.165 $\text{km}^2$/h & 0.617 $\text{km}^2$/h \\
$\alpha/\beta$ &Ratio of 1D to 2D arrival rate & 0.012 &0.012 &0.006\\
$\mu$ & Combined mortality rate & - & 0.020/h& -\\
$\mu_n$ & Nauplius mortality rate & - & - & 0.009/h\\
$\mu_c$ & Copepodite mortality rate & - & - &0.012/h\\
$\delta_m$ & Median nauplius maturation time & - & - & 251h\\
$\delta_s$ & Maturation shape parameter & - & - & 8.94\\
\hline
\end{tabular}
\label{table:paramest}
\end{sidewaystable}

\subsection{Applications}

Now that we have fit our arrival time model to the hydrodynamic simulation, we aim to answer the questions posed in the Introduction, surrounding the placement of salmon farms in a channel:
\begin{enumerate}[(i)]
\item How does the degree of cross-infection, giving by arrival probability, depend on the spacing between farms?
\item Are there scenarios where an intermediate spacing between farms leads to the highest level of cross-infection?
\item Does the relationship between cross-infection and farm spacing change in advection dominated versus diffusion dominated systems?
\item How does the maturation time for nauplii to develop into infectious copepodites affect cross-infection?
\end{enumerate}

We have a model for the distribution of arrival times of sea lice coming from a second farm, and so there are a variety of analyses that can be done with such a distribution. In this section we will focus on how various factors affect cross infection between farms, as measured by the probability of arrival, $\int_0^{\infty} e^{-\mu_c t} f_c(t)dt$.  but one could also investigate how the mean arrival time or variance of the distribution also changes. For our analyses we will use our full arrival time model that includes both maturation and survival of sea lice, as we focus on how different parameters affect the total probability that lice arrive on the farm, but for other questions relating to how the advection and diffusivity affect the arrival of particles, the simpler models may be more suitable.

To answer the above questions we explore how three different interactions of parameters affect the probability of arrival: advection ($v$) and diffusion ($D$), advection and initial farm placement ($x_0$), and median maturation time ($\delta_m$) and initial farm placement. Each of these interactions reveals a glimpse into a different component of the model, as it is difficult to gain insight if all of these parameters are changed at once. Apart from the parameters that are varying, all others will be held constant at their best fit estimates from the non-linear least squares fit, shown in Table \ref{table:paramest}. We begin by answering the last three questions before turning our attention to the first.

\subsubsection*{Are there scenarios where an intermediate spacing between farms leads to the highest level of cross-infection?}

By examining the effect of varying the advection coefficient and the placement of the first farm, we can see from Figure \ref{fig:app}a) there are indeed scenarios where an intermediate spacing leads to the highest level of cross infection. At even small advection coefficients, for example $v=0.05$, the arrival probability is maximized when the second farm is placed around $14$km away from the first. However, even if transmission to a farm 1km away may be lower than transmission to a farm 14km away, when considering self infection of a farm, local currents may allow sea lice to directly mature around the farm so that within farm infection remains high, as sea lice outbreaks on farms in the Broughton have been shown to be primarily driven by self-infection \citep{Krkosek2012}.

In fact, the probability of arrival is maximized along the line $v=mx_0+b$, for some slope $m$ and intercept $b$, and the probability decrease symmetrically as the intercept $b$ moves away from the intercept at which the probability is maximized. Intuitively this seems to be due to the relationship between the spatial mean of the solution to equation \ref{eq:pde1}, when the initial condition is a delta function. The solution is \[p(x,t)=\frac{1}{\sqrt{4\pi Dt}}e^{-\mu t-(x-x_0-vt)^2/4Dt},\] which has the spatial mean $x_0+vt$. So for a fixed maturation time, the probability of arrival should be maximized if most lice have matured before the mean density of lice moving through the channel passes by the second farm.

%The other interesting insight from Figure \ref{fig:app}b) is that depending on the advection of the channel, the arrival probability is not always maximized by having farms next to each other ($x_0=0$). At even small advection coefficients, for example $v=0.05$, the arrival probability is maximized when the second farm is placed around $14$km away from the first. However, even if transmission to a farm 1km away may be lower than transmission to a farm 14km away, when considering self infection of a farm, local currents may allow sea lice to directly mature around the farm so that within farm infection remains high, as sea lice outbreaks on farms in the Broughton have been shown to be primarily driven by self-infection \citep{Krkosek2012a}.

\subsubsection*{Does the relationship between cross-infection and farm spacing change in advection dominated versus diffusion dominated systems?}

The relationship between the advection and diffusion of the system and the arrival probability is shown in Figure \ref{fig:app}b). We can see that for any diffusion coefficient, there is a single  advection coefficient that maximizes the probability of arrival. At this maximized advection coefficient, increasing the diffusion coefficient simply reduces the probability of arrival. Now this is in the context of a fixed release location and median maturation time, but for these fixed parameters the advection coefficient plays a large role in determining whether lice will arrive at all, and the value of the diffusion coefficient determines how large the probability of arrival will be.

\subsubsection*{How does the maturation time for nauplii to develop into infectious copepodites affect cross-infection?}
 
\begin{figure}
\centering
\subfloat[]{
\includegraphics[width=7cm]{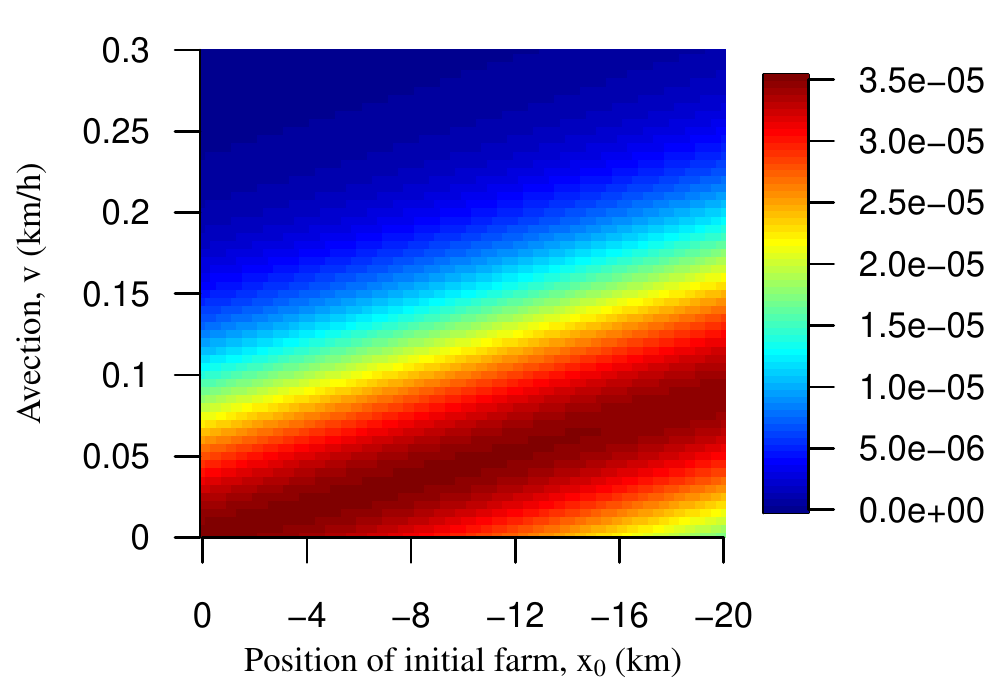}}
\subfloat[]{
\includegraphics[width=7cm]{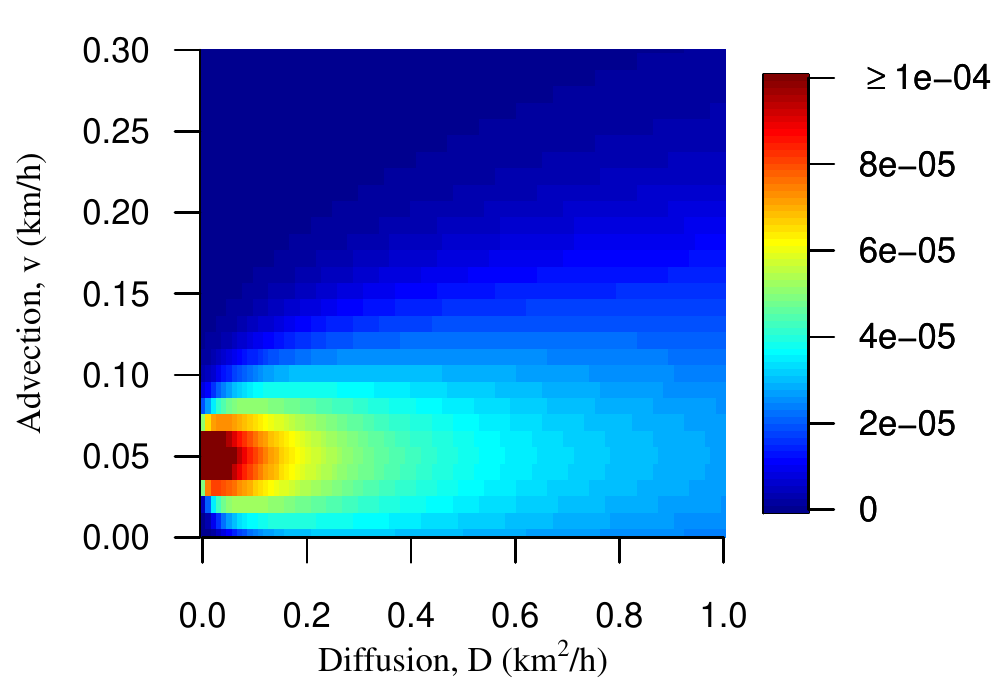}}

\subfloat[]{
\includegraphics[width=7cm]{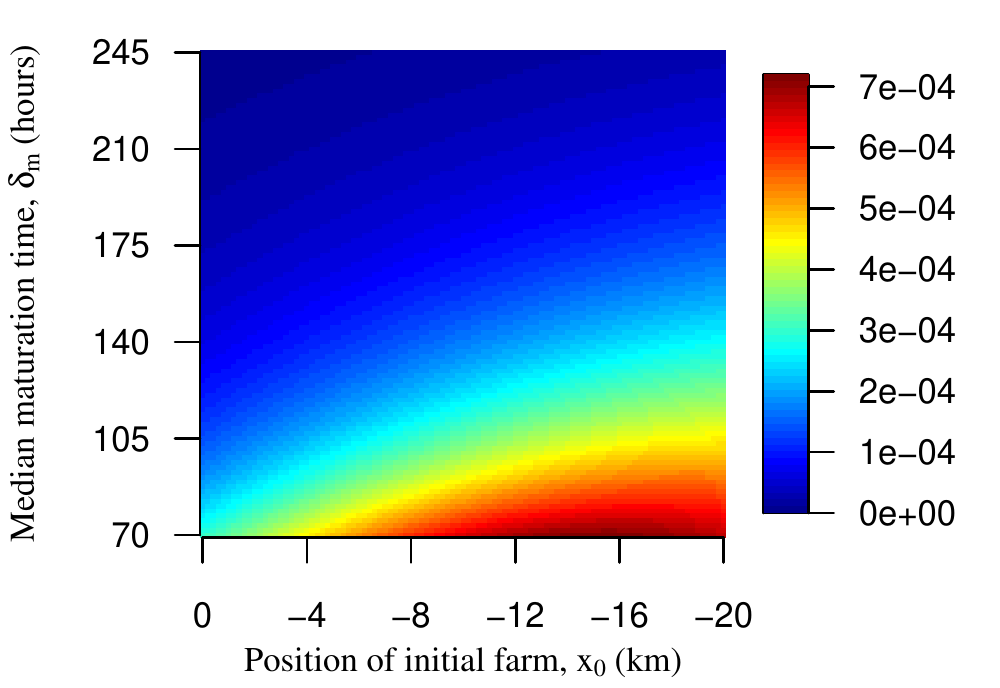}}

\caption{Total probability of sea lice arriving from one farm to another,  $\int_0^{\infty}e^{-\mu_ct}f(t)dt$, as different parameters vary: a) advection and initial farm placement, b) advection and diffusion, c) initial farm placement and median development time. Apart from the parameters that are varying, all other parameters are held constant at their best fit estimates shown in Table \ref{table:paramest}, with $\beta=1$.}
\label{fig:app}
\end{figure}

The relationship between the median maturation time, $\delta_m$, the placement of the release farm, $x_0$, and the arrival probability is shown in Figure \ref{fig:app}c). Here we chose the minimum median development time of 70 hours as this was approximately the lower end of the 95\% confidence interval for the median maturation time at 10 degrees found in a large analysis of salmon farms in Norway \citep{Aldrin2017}, and thus is most likely the fastest that nauplii would develop into copepodites in the Broughton Archipelago. Again we can see that for many median maturation times, the arrival probability is maximized at intermediate values of release farm position, $x_0$, and therefore placing farms closer to each other may not always lead to higher transmission. However for a given release farm position, the arrival probability is always maximized at the lowest possible development time. Therefore for two farms at fixed locations, warmer temperatures that cause faster louse development times will lead to higher spread between farms. 

\subsubsection*{How does the degree of cross-infection, giving by arrival probability, depend on the spacing between farms?}

We can see that the degree of cross-infection between farms depends on a variety of factors, and there is no specific farm spacing that minimizes or maximizes cross-infection across all variables. If there is very little advection in the system then cross-infection will be highest when farms are closest together. In channels with an underlying advective current, cross-infection will be maximized at some intermediate spacing, though the spacing leading to maximum cross-infection depends on the current. This is due in part to the maturation time required for sea lice to become infectious, if lice are swept by the farm before they have a chance to mature then cross-infection will be low. Complicating this relationship further is that for a given farm spacing, cross-infection may change throughout the year or among years as advection changes with river discharges and winds, as diffusion changes over spring-neap and longer tidal cycles, and as maturation time changes due to changing temperatures.

%
%Could do with drifters, temperature to parameterize dm, v, d. Now that you have simple model. 

%Calculate rough mean first passage time, along with variance. Some base quantities to observe how changing distance changes arrival.
%
%Things to consider when thinking about farm positioning:  mean time of arrival, variance, and total infection pressure, $\int_0^{\infty} f(t)dt$.
%
%How does farm placement affect these things? How does this relate to timing - figure out how this affects coordinate treatment plans with Steph's paper.
%
%Maybe also show fit to pair of farms not in channel, but rather beside each other on coastline.

\section{Discussion}

% Opening sentence of two.

The degree of sea louse connectivity between salmon farms has been well studied in specific salmon farming regions but there are few general models that can answer broad questions surrounding the effect of farm spacing and environmental variables on interfarm connectivity. In this paper we used a simple mechanistic model to calculate the timing and probabilty of arrival for sea lice dispersing between salmon farms. We then calculated the same quantities directly from complex hydrodynamical and particle tracking simulations in the Broughton Archipelago and demonstrated that our simple model captures the necessary effects of environmental and physical variables on timing and probability of arrival for sea lice in this region. Using our simple model calibrated to the Broughton Archipelago, we then investigated the effect of farm spacing, maturation time, and ocean advection and diffusion on the degree of cross-infection between farms and found that there are several scenarios in which intermediate farm spacing leads to the highest levels of cross infection.

Previous studies from the Broughton Archipelago have used hydrodynamic simulations and mechanistic models to model the dispersal of sea lice from salmon farms onto wild salmon and these studies provide useful comparisons of parameter estimates in this region (Table \ref{table:paramestcomp}). The main parameters to compare are the advection coefficient $v$, the diffusion coefficient $D$, the mortality rates $\mu_c$ and $\mu_n$.  There were no parameter estimates for the nauplius mortality rate from \citet{Krkosek2005} or \citet{Krkosek2006}, as in these studies nauplius mortality rate was estimated in conjunction with nauplius maturation rate, but most of our other parameters estimates are similar to those found previously, lending support to the accurate calibration of our simple model. 

However, the parameter values shown in table  are only estimated from one particle release day for the particle tracking model. The particular day, May 2, 2009, was the one chosen as being representative of the "average" day by \citet{Cantrell2018}. When the model is fit to a higher connectivity day (parameter estimates shown in Table \ref{table:paramesthigh} in Appendix \ref{sec:highconnectivity}), the estimate of advection (or current) speed, $v$, becomes negative and the diffusion coefficient, $D$ is much larger. In this paper our goal of fitting the simple model to hydrodynamic data is to demonstrate that the simple model can replicate realistic hydrodynamic scenarios. However if the simple model is to be used to determine the ideal placement of salmon farms in a new region, then the model should be fit to hydrodynamic data over several days during a relevant time period (the outmigration period of juvenile salmon, for example) and parameter estimates should be averaged. 

%for a given region if it is to be used to investigate the placement or removal or salmon farms, then average estimates over many representative days should be used when estimating parameters.
%
%should either be fit to particle tracks from many days, or estimates should be averaged over the outmigration period of juvenile salmon (relevant time periods, often outmigration of juvenile salmon.)

%The only parameter estimate which seems high compared to other studies is the advection coefficient $v$. This could be because in the other studies the advection coefficient is estimated over a larger region of Tribune channel than our study, and so varying coefficients in other parts of the channel could lead to a lower overall advection coefficient. In particular, \citet{Cantrell2018} found that sea lice from farms in the lower part of the channel move in the opposite direction as the advection is estimated from \citet{Krkosek2006}, which would therefore lead to a lower overall advection coefficient when estimated over the entire channel. Alternatively, our advection estimate is taken from the particle tracking simulation for a single day and may change if our model is fit against a simulation from a different time period.

In addition to confirming our parameter estimates, previous studies in the Broughton Archipelago also support our result that the highest density of farm origin sea lice in a channel may be at an intermediate distance away from the farm, leading the highest degree of cross-infection of sea lice.  In particular, \citet{Cantrell2018} found that the simulated density of infectious copepodites from the five south-easterly farms (Figure \ref{fig:maps}) was highest further than fifteen kilometers away from the nearest farm using a hydrodynamic model. Less drastically, \citet{Peacock2020} also found that the highest density of copepodites was a few kilometers away from each farm along the measured migration corridor using a mechanistic model fit to empirical data. These differing results may be a result of the different scales or methods used and are perhaps accentuated by the different maturation functions. \citet{Cantrell2018} used the same maturation process as described in this paper (equation (\ref{eq:maturation})), which can be well approximated by a Weibull function (\ref{fig:maturation}) and allows for a delay before sea lice larvae can become infections, whereas the exponential maturation function used by \citet{Peacock2020} does not allow for such a delay and thus some larvae will instantly become infectious.

When farms are not located in a channel but on nearby islands or along a common coastline the ocean dynamics will likely be diffusion dominated and thus cross infection will decrease as farms become further apart, rather than cross-infection being maximized at intermediate distances in the presence of advective currents. We found that our simple model also fits well to these scenarios (fits not shown), though in this case the absorption rate will be lower as the one dimensional approximation now effectively represents a radial slice of a two dimensional diffusion process. In these cases the probability of arrival is likely to be well-approximated by the seaway distance kernels used in other studies \citep{Aldrin2013, Aldrin2017}.

The simplicity of our mechanistic model coupled with the knowledge of how median development time changes with respect to temperature also allows us to investigate how connectivity may change as ocean temperatures warm. We found that for any fixed farm separation distance, shorter development times, which are caused by warmer temperatures, increased the probability of sea lice dispersing between farms. This result confirms previous work demonstrating that warmer temperatures increase the connectivity of farms, using a bio-physical model \citep{Cantrell2020}, and the failure to control sea louse outbreaks on wild salmon in 2015 when the water temperature was anomalously warm \citep{Bateman2016}.  Moreover, the simple analytical nature of our arrival time model and the dependence of maturation time on temperature allows us to explicitly demonstrate the dependence of connectivity on temperature, so that for the temperature of a given year, we can calculate the level of interfarm connectivity.

While our simple model can be used to understand how connectivity changes as temperatures warm, it may be beneficial to updated connectivity estimates in future years. In this case GPS drifters released from farms could be used to calculate the advection and diffusion of the ocean and determine spatial spread while average temperature and salinity data could be used to determine the appropriate maturation and survival times (equations \ref{eq:devtime}-\ref{eq:sal}). GPS drifters have previously been used to determine ocean diffusivity \citep{DeDominicis2012, Corrado2017}, and could be used to update connectivity as re-running the hydrodynamic model is computationally expensive. In British Columbia, particle tracking models with sea lice releases from farms have only been run in the Broughton Archipelago and so drifters, combined with temperature and salinity data, could be used to estimate connectivity between salmon farms in other regions across BC.

In this paper we have estimated the timing and probability of arrival using advection and diffusion estimates from particle releases that have been averaged over 24 hours to smooth the effect of the tidal cycle. However, there may be certain situations where the timing of arrival needs to be estimated at a specific point in the tidal cycle. It is also possible to calculate the arrival time distribution for this case and we present the results in Appendix \ref{sec:tidal}. We calculate the arrival time using two different methods and compare to the arrival time calculated from the complex hydrodynamical simulations with a single particle release. One method simply allows the advection coefficient to oscillate in magnitude with the tidal cycle, and the second builds on the first and also allows sea lice move between the main channel and small bays or connecting channels where they are free of the oscillating tidal flow.

Lastly, this paper has been written in the context of the current agreement between the governments of British Columbia and the Kwikwasut'inuxw Haxwa'mis, 'Namgis, and Mamalilikulla First Nations to remove salmon farms from their traditional territories \citep{Brownsey2018}. Currently 9 farms are being removed before 2023 and after 2023 7 of the remaining 11 farms will require agreements with the Kwikwasut'inuxw Haxwa'mis, 'Namgis, and Mamalilikulla First Nations and valid DFO licenses to continue to operate. Our work reinforces the notion that it is not always obvious how farm placement affects sea louse spread between farms, as there are certain instances where placing two farms further away from each other can lead to more spread than if they were closer.  We hope that our work builds on past research to help understand the level of sea louse spread between salmon farms Broughton Archipelago and that our research may help understand which farms could be the primary drivers of sea louse spread in other regions as well.

\bibliographystyle{spbasic}
\bibliography{arrivaltime.bib}

\appendix
\section{Model extension --- including tidal flow}
\label{sec:tidal}

In this paper we have modelled larval dispersal from a farm using an advection-diffusion equation, as others have done in this region \citep{Krkosek2005, Krkosek2006, Peacock2020}. When modelling dispersal using this framework, the constant advection coefficient captures  directional water movement due to river runoff and the diffusion coefficient captures the average mixing due to tidal flow and wind currents. In order to capture the average mixing due to tidal flow we have fit our arrival time model to hydrodynamic data where the spatial distribution of sea lice is drawn from 24 hours of release times. The strongest constituent tide in the Broughton is M2, which has a period of 12.4 hours, and so by sampling 24 hours of releases we capture releases from approximately two tidal periods.

\begin{figure}
\centering
\subfloat[]{
\includegraphics[width=9cm]{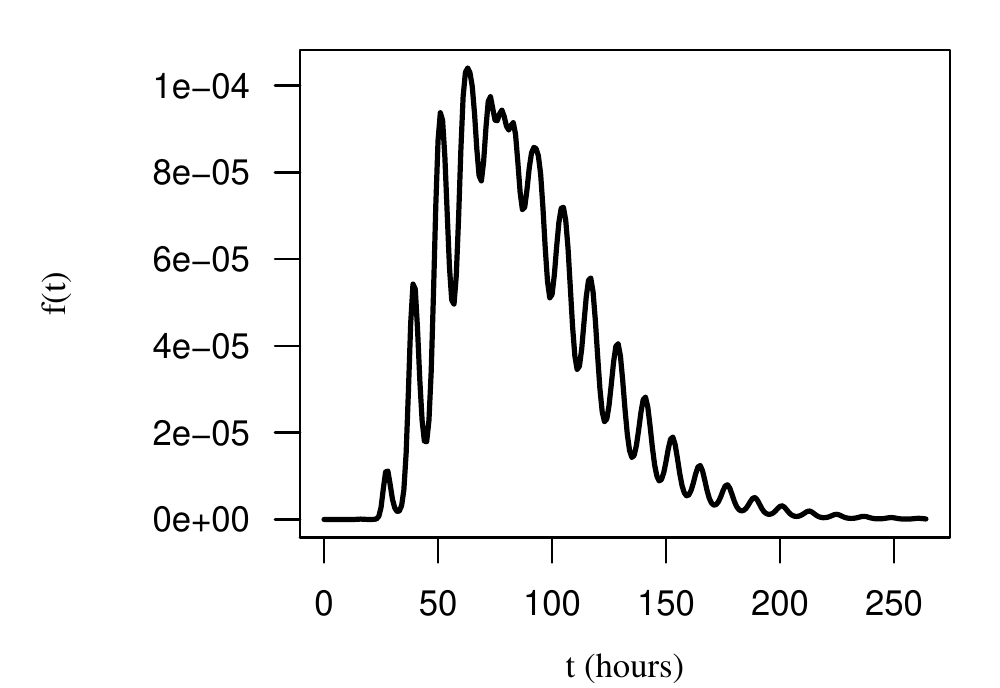}}

\subfloat[]{
\includegraphics[width=8cm]{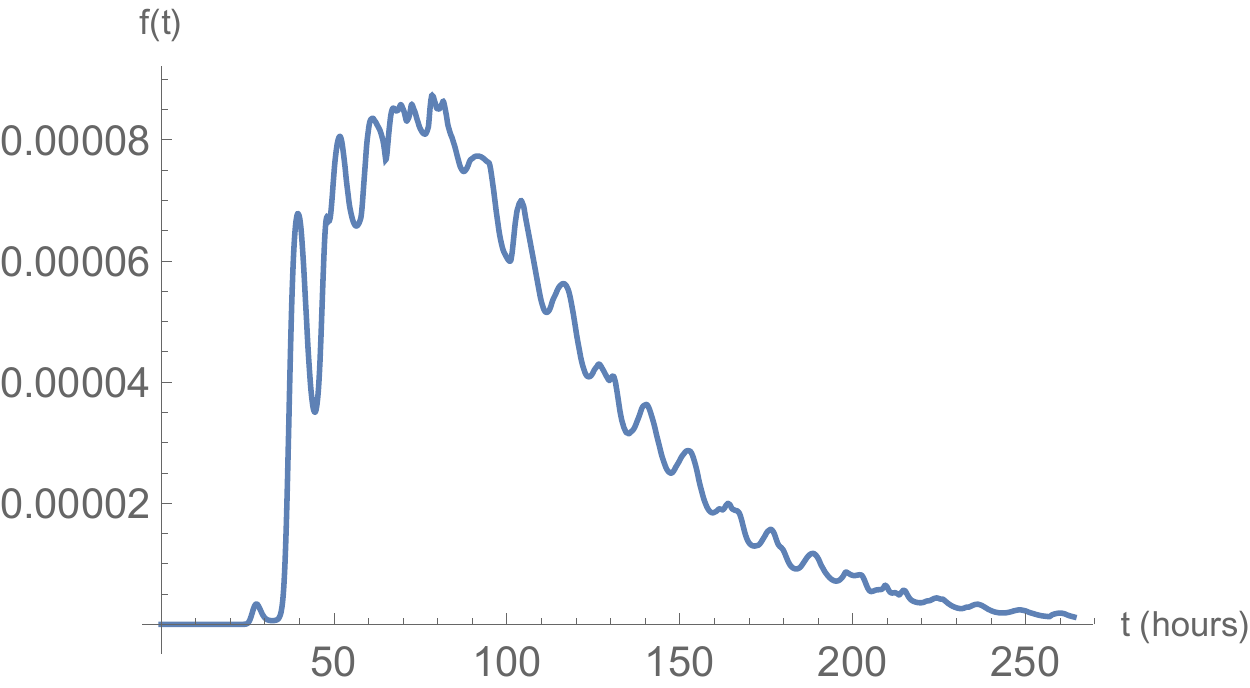}}

\subfloat[]{
\includegraphics[width=9cm]{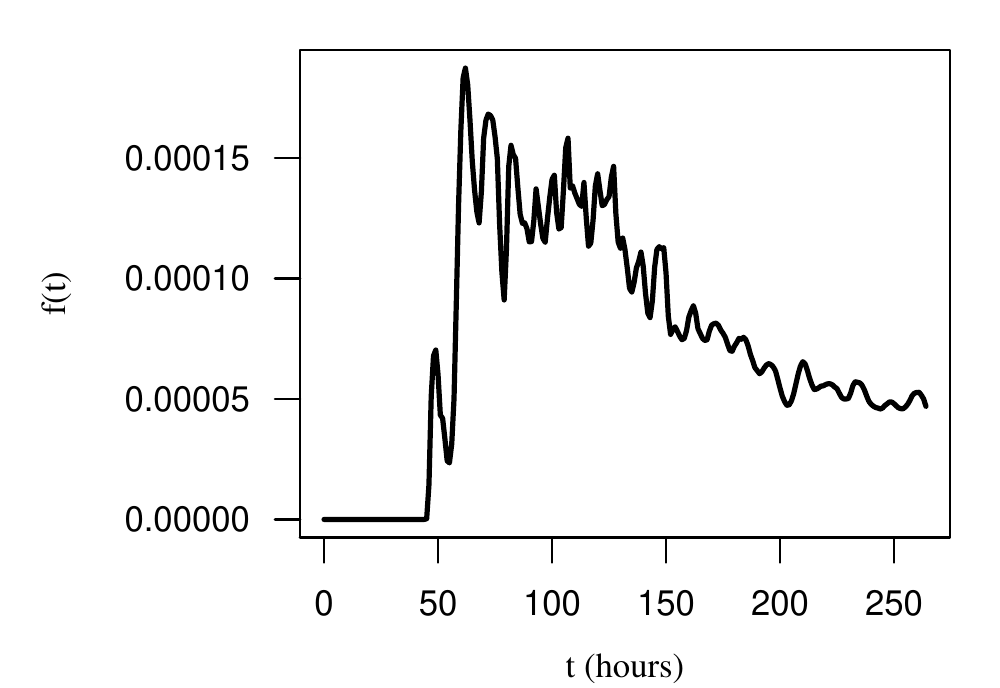}}
\caption{Arrival time distribution incorporating tidal flow. In a) the arrival time is calculated from equations \ref{eq:tide} and \ref{eq:tidef}, where the parameters are the same as the best fit parameter estimates in Table \ref{table:paramest} for the model which includes survival, with the addition of $v_1=1$, and $\beta=1$. In b) the arrival time is calculated from equation \ref{eq:tide2}, with $\alpha/\beta=0.012$, $\beta=1$, $v_0=0.3$, $v_1=1$, $D=0.01$, and $\lambda_{12}=\lambda_{21}=1/12$. In c) the arrival time is calculated directly from the KDEs and the hydrodynamic model, for a single hourly release.}
\label{fig:tidal}
\end{figure}

However, it is also possible to augment the advection-diffusion equation with constant coefficients that governs larval dispersal so that tidal flow is explicitly captured, which can then be compared to the hydrodynamic model consisting of only one release time. We briefly present two different models of larval dispersal with time varying advection coefficients to capture tidal flow and discuss their implications. Both models describe the dispersal of larvae that have survived, $\tilde{p}(x,t)$, and so their constant advection counterpart is given by equation  \ref{eq:tildep}. For simplicity we will also ignore the maturation period required for nauplii to develop into infectious copepodites before arriving onto a farm, similar to the first model presented in this paper.

The first method of modeling larval dispersal subject to tidal flow is to replace the constant advection coefficient, $v$, by an oscillating advection coefficient, $v_0+v_1 \cos(\frac{2\pi}{12}(t-t_0)$, where $v_0$ is the constant advection of the system, $v_1$ is the magnitude of the tidal flow oscillation and $t_0$ is the time in the tidal cycle at which sea lice are initially released. We assume a period of 12 hours here to demonstrate the method, though in reality the tidal period is slightly longer and would be better approximated by a summation of sinusoids each with different amplitudes, phase lags, and periods to capture the different tidal constituents. The first model for larval dispersal is then given by:

\begin{align}
\frac{\partial}{\partial t}\tilde{p}(x,t) & =-\frac{\partial}{\partial x}\left((v_0+v_1\cos(\frac{2 \pi}{12}(t-t_0)) \tilde{p}(x,t)\right)+\frac{\partial^{2}}{\partial x^{2}}\left( D\tilde{p}(x,t)\right)-h(x)\alpha \tilde{p}(x,t) \label{eq:tide}\\
\tilde{p}(x,0) & =\delta(x-x_0)\\
h(x) &= \begin{cases} 1 & x\in[0,L]\\
0 & \text{otherwise}
\end{cases}.
\end{align}
The benefit of this framework is that it is again possible to non-dimensionalize, perform an asymptotic expansion in the small parameter, $\epsilon=\alpha L^2/D$, solve for the arrival time distribution, $f(t)$, up to order $\epsilon^2$, and then re-dimensionalize $f(t)$ which is then given by:
\begin{align}
f(t)&=\frac{\alpha}{2}\left(\erf \left( \frac{x_0+v_0 t+v_1  \frac{12}{2 \pi }\left(\sin\left(\frac{2\pi }{12 }(t-t_0)\right)+\sin\left(\frac{2\pi }{12 }(t_0)\right) \right)}{\sqrt{4Dt}} \right) \right.\\
&\left. -\erf \left(\frac{x_0+v_0 t+ v_1 \frac{12}{2 \pi }\left(\sin\left(\frac{2\pi}{12}(t-t_0)\right)+\sin\left(\frac{2\pi }{12 }(t_0)\right) \right) -L}{\sqrt{4Dt}} \right) \right)\label{eq:tidef},
\end{align}
shown in Figure \ref{fig:tidal}a). The downside to this approach is that the average over all tidal release times, $t_0$, is simply the constant advection diffusion equation with the same diffusion coefficient as the oscillating advection case. This seems somewhat biologically unrealistic, as the addition of modeling tidal flow was meant to allow for a smaller diffusion coefficient, as the mixing due to tidal flow should now be accounted for in the oscillating advection term.

In order to construct more realistic equations where the oscillating tidal flow mimics the diffusion in the constant advection diffusion equation, for the second model we split the ocean environment into two compartments, a lateral compartment that represents movement along the channel, and a cross channel compartment where larvae remain stationary in the lateral direction. This cross channel compartment represents larvae that may be swept into eddies, small bays along the shore, or into any connecting channels that are perpendicular to the lateral channel. We let $\tilde{p}(x,t)$ denote the larvae that are in the lateral compartment, $q(x,t)$ to be the larvae that are in the cross channel compartment, $\lambda_1$ to be the rate of transfer from the lateral to cross channel compartment, and $\lambda_2$ to be the rate of transfer between the cross channel and lateral compartment. The model is

\begin{align}
\frac{\partial}{\partial t}\tilde{p}(x,t) & =-\frac{\partial}{\partial x}\left((v_0+v_1\cos(\frac{2 \pi}{12}(t-t_0)) \tilde{p}(x,t)\right)+\frac{\partial^{2}}{\partial x^{2}}\left( D\tilde{p}(x,t)\right)-h(x)\alpha \tilde{p}(x,t) \nonumber\\
&-\lambda_1 \tilde{p}(t,x)+\lambda_2 q(t,x) \label{eq:tide2}\\
\frac{d}{dt}q(x,t)&=\lambda_1 \tilde{p}(x,t)-\lambda_2 q(x,t)\\
\tilde{p}(x,0) & =\delta(x-x_0)\\
q(x,0)&=0\\
h(x) &= \begin{cases} 1 & x\in[0,L]\\
0 & \text{otherwise}
\end{cases}.
\end{align}

With the addition of this cross channel compartment, where larvae can remain stationary along the channel, this model can replicate similar arrival times as the previous model but with much smaller diffusion coefficients. The cross channel compartment allows some larvae to be swept forward in the lateral compartment, transfer to the cross channel compartment and remain there while other larvae may be swept back, before re-entering the lateral compartment and moving forward. The only downside to this model is it is no longer simple to perform an asymptotic analysis and arrive at an analytical expression for the arrival time.

To demonstrate that this model can replicate a similar arrival time distribution as the previous model, we compare the arrival time distributions for the two models and the arrival time calculated from a single louse release in the hydrodynamic mode in Figure \ref{fig:tidal}. For this figure we set $\lambda_1=\lambda_2=1/12$, so that the average time spent in either compartment was 12 hours, the same as the tidal period. Here we can see that the two arrival time distributions are similar, but the arrival time from the two compartment model has a much lower diffusion coefficient.

%A comparison of this model, the previous model, and 
%
%
%Here include model with oscillating tidal flow, as well as oscillating tidal flow with lateral and cross channel compartments.
%
%For tidal flow, just provide equations and maybe solution? For lateral flow just provide equations. Talk about differences in v and D when cross channel compartment is added. Different effective diffusion, and parameter meanings. Show figure with a) tidal flow results, b) lateral flow results c) actual data (pick good one). 
%
%Model is robust to differences in formulation, but parameter interpretations may change. 

\section{Arrival time fit to high connectivity day}
\label{sec:highconnectivity}

\begin{figure}
\centering
\subfloat[]{
\includegraphics[width=7.5cm]{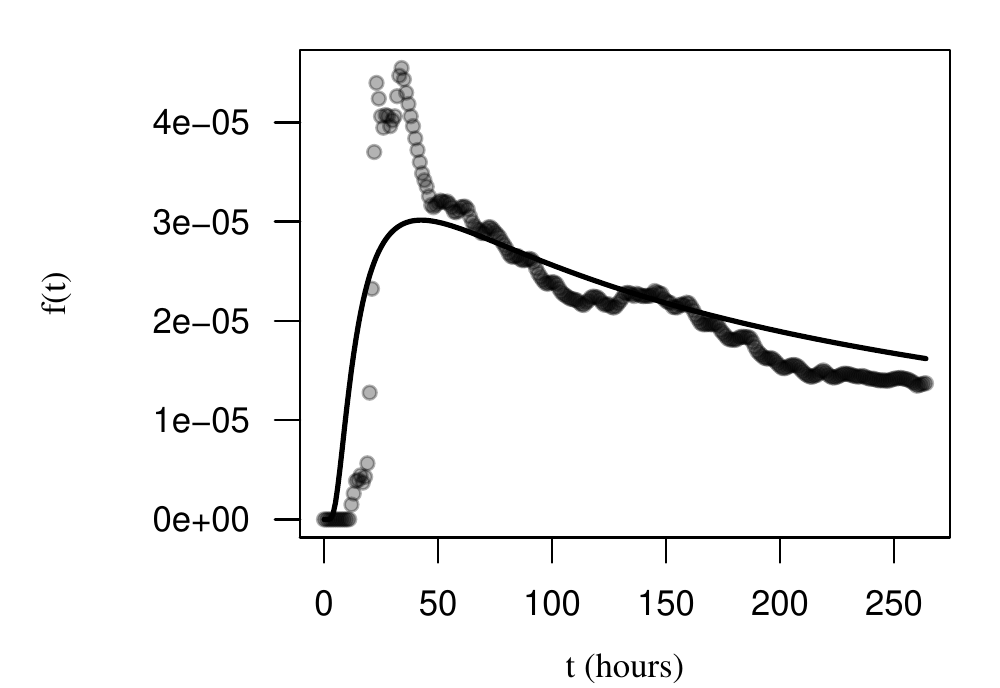}}
\subfloat[]{
\includegraphics[width=7.5cm]{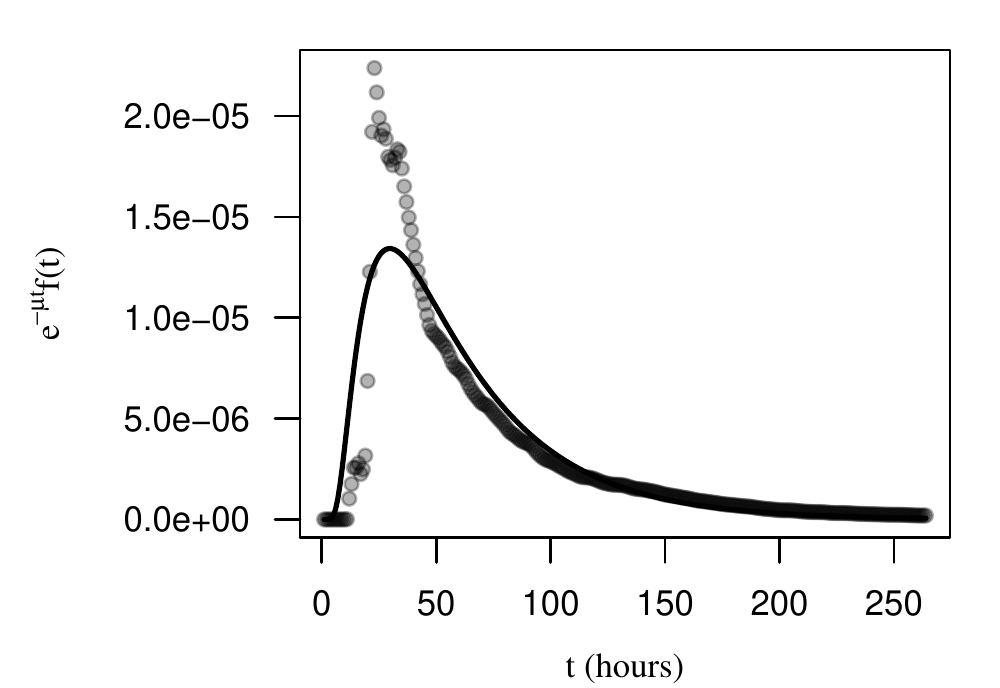}}

\subfloat[]{
\includegraphics[width=7.5cm]{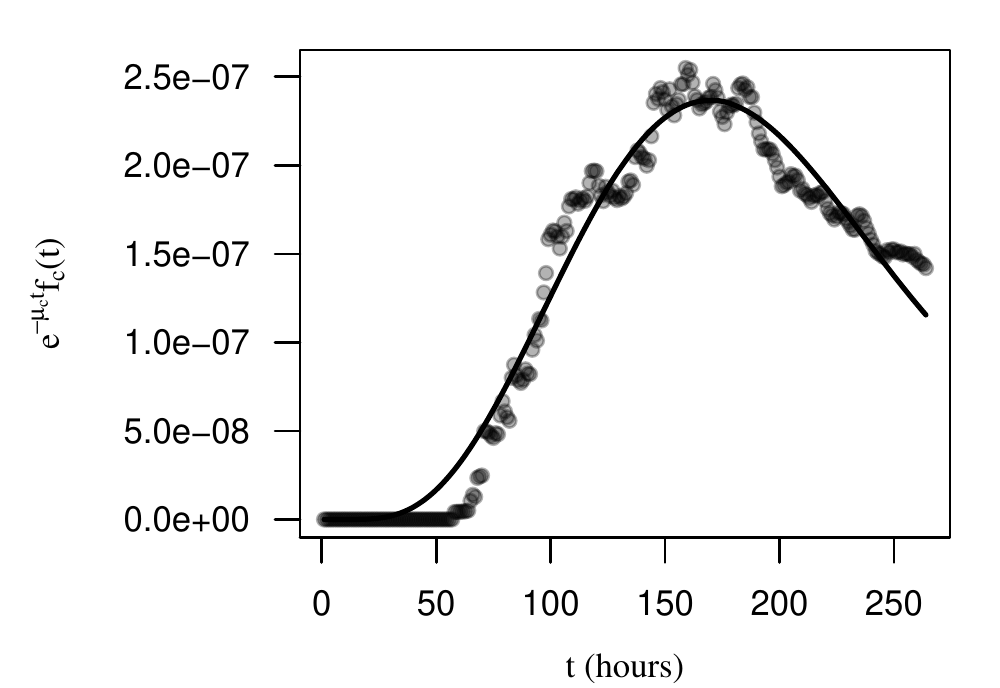}}

\caption{Arrival time distribution of a) inert particles, b) particles with survival included, and c) infectious copepodites, calculated from a high connectivity day in the hydrodynamic model (May 11, 2009). In a) the points are the arrival time densities calculated from the KDEs of the particle locations, in b) the points are the arrival time densities calculated from the KDEs of particle locations weighted by survival, and in c) the points are the arrival time densities calculated from infectious particle locations weighted by survival. The curves are the best fit lines of the corresponding analytical models fit to these points via non-linear least squares. Parameter estimates for the model are given in Table \ref{table:paramesthigh}, with $\beta=1$.}
\label{fig:fhigh}
\end{figure}

Here we present the parameter estimates (Table \ref{table:paramesthigh}) and model fits (Figure \ref{fig:fhigh})  for the arrival time distributions fit to particle tracking results of a high connectivity day (May 11, 2009 or CRD 59 in \citet{Cantrell2018}).

\begin{sidewaystable}
\centering
\caption{Parameter estimates of best fit models under non-linear least squares for a high connectivity day.}

\begin{tabular}{c l l l l}
\hline
Parameter & Description & Inert & With survival & With survival and maturation\\
\hline
$v$ & Advection & -0.075 km/h & -0.055 km/h & -0.057 km/h\\
$D$ & Diffusion & 2.035 $\text{km}^2$/h & 1.317 $\text{km}^2$/h & 0.647 $\text{km}^2$/h \\
$\alpha/\beta$ &Ratio of 1D to 2D arrival rate & 0.011 &0.0125 &0.005\\
$\mu$ & Combined mortality rate & - & 0.022/h& -\\
$\mu_n$ & Nauplius mortality rate & - & - & 0.013/h\\
$\mu_c$ & Copepodite mortality rate & - & - &0.012/h\\
$\delta_m$ & Median nauplius maturation time & - & - & 171h\\
$\delta_s$ & Maturation shape parameter & - & - & 3.44\\
\hline
\end{tabular}
\label{table:paramesthigh}
\end{sidewaystable}

%
%\begin{figure}
%\centering
%\includegraphics[width=9cm]{paperplotcrd59.pdf}
%\caption{Arrival time distribution of inert particles. The points are the arrival time densities calculated from the hydrodynamic model on a high connectivity day (May 11, 2009) and the curve is the best fit line of the simple analytical model fit to these points via non-linear least squares. Parameter estimates for the model are given in Table \ref{table:paramesthigh}, with $\beta=1$.}
%\label{fig:fhigh}
%\end{figure}
%
%
%
%\begin{figure}
%\centering
%\includegraphics[width=9cm]{paperplotcrd592.pdf}
%\caption{Arrival time distribution of particles with survival included. The points are the arrival time densities calculated from the hydrodynamic model on a high connectivity day (May 11, 2009) with KDE estimates weighted by survival and the curve is the best fit line of the simple analytical model fit to these points via non-linear least squares. Parameter estimates for the model are given in Table \ref{table:paramesthigh}, with $\beta=1$.}
%\label{fig:fshigh}
%\end{figure}
%
%
%\begin{figure}
%\centering
%\includegraphics[width=9cm]{paperplotcrd593.pdf}
%\caption{Arrival time distribution of infectious copepodites. The points are the arrival time densities calculated from the hydrodynamic model on a high connectivity day (May 11, 2009) and the curve is the best fit line of the simple analytical model fit to these points via non-linear least squares. Parameter estimates for the model are given in Table \ref{table:paramesthigh}, with $\beta=1$.}
%\label{fig:fsmhigh}
%\end{figure}

\end{document}